\documentclass{article}

\usepackage{textcomp}

\usepackage{amsmath,amssymb}

\usepackage{bm}
\usepackage{psfrag}
\usepackage{epsfig}
\usepackage{graphicx,subfigure}
\usepackage{mathrsfs}
\usepackage{url}
\usepackage{bbm,color}
\usepackage{pifont}
\usepackage[title]{appendix}
 \usepackage{booktabs}
\usepackage{algorithm}
\usepackage{algpseudocode}
\usepackage{tikz}
 \newtheorem{assumption}{Assumption}
 
 \newtheorem{theorem}{\textbf{Theorem}}
 \newtheorem{proposition}{\textbf{Proposition}}
 
\newtheorem{remark}{\textbf{Remark}}
 \newtheorem{example}{\textbf{Example}}

\newtheorem{definition}{Definition}

\def\ba{\begin{aligned}}
\def\ea{\end{aligned}}

\newcommand{\mV}{\mathrm{V}}
\newcommand{\mG}{\mathrm{G}}
\newcommand{\mE}{\mathrm{E}}

\newcommand{\qb}{\mathbf{q}}

\newcommand{\vb}{\mathbf{v}}

\newcommand{\xb}{\mathbf{x}}
\newcommand{\yb}{\mathbf{y}}
\newcommand{\zb}{\mathbf{z}}

\newcommand{\Hb}{\mathbf{H}}
\newcommand{\Lb}{\mathbf{L}}
\newcommand{\Ib}{\mathbf{I}}

\newcommand{\Fb}{\mathbf{F}}
\newcommand{\Eb}{\mathbb{E}}

\newcommand{\Cb}{\mathcal{C}}

\newcommand{\calSo}{\mathbf{s}}

\newcommand{\Fl}{\tilde{\Fb}}

\newcommand{\xl}{\overline{\xb}}
\newcommand{\xhl}{\overline{\xb}_c}
\newcommand{\vl}{\overline{\vb}}

\newcommand{\psib}{\bm{\psi}}

\newcommand{\xch}{\mathbf{\overline{x}}_{\perp}}
\newcommand{\xpi}{\mathbf{\overline{x}}_{\parallel}}
\newcommand{\vch}{\mathbf{\overline{v}}_{\perp}}
\newcommand{\vpi}{\mathbf{\overline{v}}_{\parallel}}
\newcommand{\xchc}{\mathbf{\overline{x}}_{\perp}}
\newcommand{\xpic}{\mathbf{\overline{x}}_{\parallel}}
\newcommand{\vchc}{\mathbf{\overline{v}}_{\perp}}
\newcommand{\vpic}{\mathbf{\overline{v}}_{\parallel}}

\newcommand{\sigmab}{\boldsymbol{\sigma}}

\newcommand{\calSb}{\mathbf{S}_\otimes}
\newcommand{\onb}{\mathbf{1}_\otimes}

\begin{document}

\title{Distributed Optimization by Network Flows with Spatio-Temporal Compression (Extended Version)} 
\author{
Zihao Ren, Lei Wang, Xinlei Yi, Xi Wang, Deming Yuan, Tao Yang,\\ Zhengguang Wu, Guodong Shi
\thanks{A preliminary version of the paper  {was} presented at the 63rd IEEE Conference on Decision and Control,  December 16-19, 2024, Allianz MiCo, Milan Convention Centre, Italy \cite{ZR-ST}. {\em (Corresponding author: Lei Wang)}
}
\thanks{Z. Ren, L. Wang and Z. Wu are with State Key Laboratory of Industrial Control Technology, Institute of Cyber-Systems and Control, Zhejiang University, Hangzhou 310027, China. (E-mail:  \{zhren2000; lei.wangzju; 
 nashwzhg\}@zju.edu.cn)
}
\thanks{X. Yi is with the College of Electronics and Information Engineering, Tongji University, China. (E-mail:xinleiyi@tongji.edu.cn)}
\thanks{X. Wang is with the School of Electrical Engineering  Telecommunications,
University of New South Wales, Sydney,  Australia. (E-mail: xi.wang14@unsw.edu.au)}
\thanks{D. Yuan is with the School of Automation, Nanjing University of Science and Technology,  China. (E-mail: dmyuan1012@gmail.com)}
\thanks{T. Yang is with the State Key Laboratory of Synthetical Automation for Process Industries, Northeastern University,  China. (E-mail: yangtao@mail.neu.edu.cn)}
\thanks{G. Shi is with the Australian Centre for Robotics, School of Aerospace, Mechanical and Mechatronic Engineering, The University of Sydney, Sydney, NSW 2006, Australia. (E-mail:guodong.shi@sydney.edu.au)}
}

\maketitle

\begin{abstract}
Several data compressors have been proposed in distributed optimization frameworks of network systems to reduce communication overhead in large-scale applications. In this paper, we demonstrate that effective information compression may occur over time or space during sequences of node communications in distributed algorithms, leading to the concept of spatio-temporal compressors. This abstraction classifies existing compressors  {and inspires new compressors} as spatio-temporal compressors, with their effectiveness described by constructive stability criteria from nonlinear system theory.
Subsequently, we incorporate these spatio-temporal compressors  {\emph{directly}} into standard continuous-time consensus flows and distributed primal-dual flows, establishing conditions ensuring exponential convergence. Additionally, we introduce a novel observer-based distributed primal-dual continuous flow integrated with spatio-temporal compressors, which provides \emph{broader} convergence conditions. These continuous flows achieve exponential convergence to the global optimum when the objective function is strongly convex and can be discretized using Euler approximations.
Finally, numerical simulations illustrate the versatility of the proposed spatio-temporal compressors and verify the convergence of algorithms.
\end{abstract}


\section{Introduction}
Distributed intelligent systems, such as drone swarms, smart grids, and cyber-physical systems, have been extensively researched across disciplines such as control, signal processing, and machine learning \cite{magnusbook}-\cite{Rabbat2010}. The mathematical representation of a distributed system involves a network connecting multiple agents, where each node symbolizes an individual agent, and the edges depict communication lines between these nodes. 
When distributed systems are required to implement tasks such as cluster optimization and collaborative control, it is required to compute distributively. In this process, each node stores localized information, communicates messages with connected nodes through the network, and collaboratively solves a mathematical problem \cite{magnusbook}. This paper focuses on addressing distributed optimization problems, where each node possesses a function, aiming to identify solutions that collectively minimize the sum of all functions through communication across the network. 

An extensive effort has been devoted to developing distributed optimization algorithms based on the consensus algorithm. The goal of the consensus algorithm is to reach a consensus of the states across nodes. A combination of the consensus algorithm with the classical gradient descent method in optimization problems, coupled with stability tactics, results in the distributed (sub)gradient algorithm (DSG), achieving sublinear convergence under a strongly convex global objective function \cite{SMAC}-\cite{Shi-eq}.
More algorithms have been introduced to address distributed optimization problems with faster rate requirements,  {e.g., linear convergence}. For example, the distributed gradient tracking algorithm (DGT) incorporates an additional state to track the gradient of the objective function \cite{PGMF}, akin to integral action \cite{TGTI}. 
Besides, for different formulations of distributed optimization, various Lagrangian functions have been proposed, giving rise to multiple algorithms based on the saddle point dynamic method. Examples include the Wang--Eila algorithm in \cite{JW-WE} and the primal-dual algorithm in \cite{XY-LCOF}, distinct in  communication states.

In practical implementation, the network bandwidth for communication in distributed systems is limited and numerous strategies have been developed to address this issue. In  \cite{evtr}, an event-triggered communication strategy is proposed to reduce communication frequency, thereby alleviating the communication burden. In addition, compressors that reduce the  communication burden in each round have been extensively studied.
Specifically, several compressors capable of reducing communication bits are proposed by synthesizing concepts from quantization \cite{CROD}-\cite{XY-CCFD}, sparsity \cite{ALCA}, scalarization \cite{LW-DSFN} and randomization \cite{DSOA,AR-AEQD}. Instead of specific ones, there are also some results that propose a general class of compressors \cite{ICFC}-\cite{CGMW}, which contains some existing specific compressors and also allows to explore new compressors.
 {
However, the compressor classes proposed in the literature mainly focus on the spatial dimension, utilizing the information contained within transmitted messages. This prevents new compressors that utilize temporal information (e.g., \cite{LW-DSFN}) from being included, which motivates our study of new compressor classes that capture both temporal and spatial information.
}

In addition,  how to combine the compressors with distributed optimization algorithms has become a noteworthy area of study. This is because the compression method can facilitate the successful integration of more general compressors and enhance the effectiveness of the algorithm. 
Direct compression often leads to biased convergence \cite{DSMA,QIAF}. To handle this, more complex compression methods with extra states are proposed. For instance, \cite{CROD,AR-AEQD} incorporate a weighted sum of the updated value and the original value into the original value, while \cite{FCRO,DACW} compress the difference between iterations rather than the original value. In \cite{LCOC,DCWL}, the difference is scaled and then compressed, with the results communicated after a reverse reduction  {to ensure convergence}. In
\cite{XY-CCFD,ACGT,CGTA,liu2021linear} a difference compression method based on filters is adopted,  {where only compressed values are exchanged through some additional equivalent transformations.
Despite these impressive results, it is still open how to achieve unbiased linear convergence when directly incorporating the compressors into distributed optimization algorithms.
This then acts as part of our research motivations.}

 {In view of the previous analysis, this paper aims to propose a new and general compressor class characterized by properties of simultaneously capturing both temporal and spatial dimensions, i.e., the spatial-temporal (ST) compressor. Given this compressor class and considering the fact that the majority of existing distributed optimization algorithms are established on the consensus, we first investigate the condition under which direct compression in the consensus flow can lead to unbiased exponential convergence. To incorporate more ST compressors into algorithms, we further propose a novel observer-based compression method. With these compressed consensus flows, we then propose two distributed ST-compressed primal-dual flows with direct  and observer-based compression, respectively, for which exponential convergence can be guaranteed. Moreover, for implementation, discrete-time algorithms are  established by Euler discretization with linear convergence guarantees.
The main contribution  lies in the following aspects.
\begin{itemize}
    \item A general class of ST compressors and its strong version are proposed for communication compression in distributed optimization, from a novel perspective of nonlinear system theory, specifically, in terms of exponential stability of nonlinear non-autonomous systems. This ST compressor class not only encompasses various existing communication compressors, but also inspires new compressors. 
    \item An affirmative answer is established that \emph{directly} incorporating the strong ST compressors can lead to unbiased exponential/linear convergence, if a sufficient condition is satisfied.  It is also shown that such a condition is satisfied if the compressor is linear, e.g., the scalarization compressor.
    \item An observer-based compression method is developed so that more general ST compressors can be incorporated into distributed algorithms with exponential/linear convergence guaranteed, while removing the restrictive sufficient condition for the direct compression method. 
\end{itemize}}

 We begin with and focus on continuous-time systems in this paper, even though distributed optimization algorithms in real applications are mostly in discrete-time. This is because, first, the properties of our ST compressors are described by the induced systems, and the converse Lyapunov theorem is used, with the continuous-time form being more intuitive for our analysis. Second, we would like to build our work as a natural extension of the results on continuous-time distributed optimization algorithms in the control system community, e.g., \cite{cort}-\cite{Shao}. In addition, we offer a discrete version of the proposed continuous algorithm by Euler approximation,    enabling our results to be applied in real situations. 

The paper is structured as follows. Section \ref{sec.pro} formulates the distributed optimization problem and proposes the notion of (strong) spatio-temporal compressor for message communication. In Section \ref{sec.ave}, we start from the distributed consensus to illustrate the conditions required by the direct compression method, and then introduce the  observer-based compression method. In Section \ref{sec.CPD}, we respectively discuss the applicability of these two compression methods to the primal-dual flow.
In Section \ref{sec.DPD}, we propose the ST compressors in discrete time and discretize the continuous-time flows by the Euler method.  Numerical simulations are presented to show the effectiveness of the proposed algorithms in Section \ref{sec.num}. Finally, a conclusion is made in Section \ref{sec.con}.  All proofs are collected in Appendices. 
 {Compared to our preliminary conference version \cite{ZR-ST}, this paper has made several new significant results, by introducing the concept of strong ST compressors and proposing distributed compressed consensus/optimization flows with direct  and observer-based   compression methods (see Theorems 1-4). Furthermore, we introduce the ST compressors in discrete time and propose two distributed compressed optimization algorithms via Euler discretization.}

\emph{\bf Notations.}
$\left\|\cdot \right\|$
denotes the Euclidean norm. The notation $\mathbf{1}_n\left(\mathbf{0}_n\right)$, $\mathbf{I}_n$ and $\left\{{\bf e}_1,\dots,{\bf e}_n\right\}$ denote column one (zero vector, identity matrix and base vectors in $\mathbb{R}^n$, respectively. Denote $\mathrm{diag}\left(x_1,\dots,x_n\right)$ as a
diagonal matrix with the $i$-th diagonal element being $x_i$. The symbol $\otimes$ denotes the Kronecker product. We use $\nabla\left(\cdot\right)$ to denote the gradient of a function and use $*$ to denote the Hadamard product. 

\section{Problem Formulation}
\label{sec.pro}

\subsection{Distributed Optimization}

 Consider a network of agents indexed by $\mathrm{V}=\left\{1,2,\dots,n\right\}$, where each agent $i\in\mathrm{V}$ holds an objective function $f_i:\mathbb{R}^d\rightarrow \mathbb{R}$. The agents aim to solve the following system-level optimization problem 
\begin{equation}
    \label{eq:DO}
    \ba
    \mathrm{min}&\  \sum_{i=1}^n f_i\left(\xb_i\right),\\
    \mathrm{s.t.}&\ \xb_i=\xb_j, \quad\forall i,j\in \mV.
    \ea
\end{equation}
In particular, each local objective function $f_i$ is assumed to satisfy the following requirements.
\begin{assumption}\label{ass-DO}
    The following properties are satisfied.
    \begin{itemize}
    \item[i).] The \emph{global} objective function $f\left(\xb_e\right):=\sum_{i=1}^n f_i\left(\xb_e\right)$ is strongly convex, i.e., there exists $\mu>0$ such that 
    $f\left(\xb'_e\right)\geq f\left(\xb_e\right)+\nabla f\left(\xb_e\right)^T\left(\xb'_e-\xb_e\right)+\frac{\mu}{2}\left\|\xb'_e-\xb_e\right\|^2$ for all $\xb'_e,\xb_e\in\mathbb{R}^d$.   
    \item[ii).] Each local gradient  $\nabla f_i$ is globally Lipschitz continuous, i.e., there exists $L_f>0$ such that  for all $\xb_e,\xb'_e\in\mathbb{R}^d$, $\left\|\nabla f_i\left(\xb_e\right)-\nabla f_i\left(\xb'_e\right)\right\| \leq L_f\left\|\xb_e-\xb'_e\right\|$. \hfill$\square$ 
    \end{itemize}
\end{assumption}

If Assumption \ref{ass-DO} holds, then the considered optimization problem \eqref{eq:DO} turns out a strongly convex optimization problem, allowing a unique solution  {$s^\ast{\in}\mathbb{R}^d$} such that 
$\nabla f\left(s^\ast\right)=0$ and $f\left(s^\ast\right)=f^\ast$, where $f^\ast$ is the optimal value.

As each agent only has information about its own objective function, to solve such a distributed optimization problem \eqref{eq:DO}, a communication network is usually required to transmit messages. Denote the communication graph $\mathrm{G=\left(V,E\right)}$, where $\mathrm E$ denotes the set of edges. 
Let $[a_{ij}]\in \mathrm{R}^{n\times n}$ denote the weight matrix, i.e., $a_{ij}>0$ if $\left\{j,i\right\}\in\mE$ and $a_{ij}=0$ if $\left\{j,i\right\}\notin\mE$. Then denote the Laplacian matrix of graph $\mathrm G$ by $\Lb$, satisfying $[\Lb]_{ij}=-a_{ij}$ for all $i\neq j$, and $[\Lb]_{ii}=\sum_{j=1}^n a_{ij}$ for all $i\in\mathrm{V}$. Denote the neighbor set of agent $i$ as $\mathrm{N}_i$, satisfying $j\in\mathrm{N}_i$ if and only if $[\Lb]_{ij}\neq0$ for all $i,j\in\mV$. For simplicity, we make the following assumption on the communication graph.

\begin{assumption}\label{ass-graph}
    The graph $\mG$ is undirected and connected.
\end{assumption}

Assumption \ref{ass-graph} indicates that the Laplacian matrix $\Lb$ is symmetric and positive semi-definite, with $[\Lb]_{ij}=[\Lb]_{ji}$, $ \Lb\mathbf{1}_n=\mathbf{0}_{n}$ and its eigenvalues $\lambda_i$, $i\in\mV$ in ascending order satisfying $0=\lambda_1<\lambda_2\leq\dots\leq \lambda_n$  by \cite{magnusbook}. We let $\mathbf{S}\in\mathbb{R}^{n\times\left(n-1\right)}$ be a matrix whose rows are eigenvectors corresponding to non-zero eigenvalues of 
$\Lb$,  satisfying
\[
\ba 
\mathbf{S}^T\mathbf{1}_n =\mathbf{0}_{n-1}\,,\quad  \mathbf{I}_n = \mathbf{S}\mathbf{S}^T +\mathbf{1}_n\mathbf{1}_n^T/n .
\ea
\]
In the literature, various distributed optimization algorithms have been developed to compute the solution $s^\ast$ for \eqref{eq:DO} \cite{PGMF}-\cite{XY-LCOF}. In this paper, we  focus on the distributed primal-dual algorithm, which enables exponential convergence and further generalizations to the case with constraints \cite{DJ-PD1,PC-PD2}\footnote{ {Though only distributed primal-dual algorithm is considered, we stress that the proposed compressors and compression methods in this paper can be incorporated into other common consensus-based algorithms, e.g. DGT in \cite{PGMF} or the Wang-Eila algorithm in \cite{JW-WE}.}}. A common distributed primal-dual flow for \eqref{eq:DO} takes the form \cite{JW-WE,XY-LCOF,S-Kia}
\begin{equation}
\ba \label{eq:Primal_Dual}
\dot{\xb}_i=-\sum^{n}_{j=1}\Lb_{ij} {\xb}_{j}-\beta \vb_i-\eta \nabla f_i\left(\xb_i\right), \
\dot{\vb}_i={\beta\sum^{n}_{j=1}\Lb_{ij}\xb_{j}},
\ea
\end{equation}
where $\beta,\eta>0$ are parameters to be fixed and the initial condition  $\sum_{i=1}^n \vb_i\left(0\right)=\mathbf{0}_d$.

\vspace{-0.2cm}
\subsection{Spatio-Temporal Compressors}

We propose the following notions of (strong) spatio-temporal compressors for compressing node-to-node communications in distributed algorithms.



 {
\begin{definition}[Spatio-Temporal  Compressor]
\label{def-SST}
The mapping $\mathbf{C}:\mathbb{R}^d\times\mathbb{R}_+\rightarrow \mathbb{R}^d$ is said to be a \emph{spatio-temporal (ST) compressor}, if the following two properties hold.
\begin{itemize}
    \item[P1).] There exists a $k>0$ such that the induced continuous-time non-autonomous system $\dot\xb_e=-k\mathbf{C}\left(\xb_e,t\right)$ is uniformly globally exponentially stable (UGES) at the origin;
    \item[P2).] There exists a $L_c>0$ such that 
    \begin{equation}\label{eq:ugl}
        \left\|\mathbf{C}\left(\xb_e,t\right)-\mathbf{C}\left(\xb_e',t\right)\right\| \leq L_c\left\|\xb_e-\xb_e'\right\|
    \end{equation} 
    for all $\xb_e\in\mathbb{R}^d,\xb_e'=0$ and any $t\in\mathbb{R}_+$. 
\end{itemize} 
Such mapping $\mathbf{C}$ is said to be a \emph{strong spatio-temporal (SST) compressor}, if the UGES property in P1) holds for all $k>0$, and  \eqref{eq:ugl} in P2) holds for all $\xb_e,\xb_e'\in\mathbb{R}^d$ and any $t$, namely, the mapping $\mathbf{C}$ is \emph{uniformly globally Lipschitz}.
\hfill$\square$
\end{definition}}

For a ST compressor $\mathbf{C}$, it needs to vanish at the origin, i.e., $\mathbf{C}\left(0,t\right)\equiv0$ uniformly in $t$, to satisfy the UGES property. This immediately implies that P2) turns out the \emph{uniformly linearly bounded} property, i.e., $\left\|\mathbf{C}\left(\xb_e,t\right)\right\| \leq L_c\left\|\xb_e\right\|$.  {In addition,  it is clear that  $k\mathbf{C}$ is also a ST compressor if so is the mapping $\mathbf{C}$ by definition. Thus, we can simply incorporate such $k$ when designing compressors. In view of this, without loss of generality,  we assume UGES of $\dot{\xb}_e=-\mathbf{C}\left(\xb_e,t\right)$  by letting $k=1$ for simplicity, when referring to a ST compressor $\mathbf{C}$ in the sequel. By the converse Lyapunov Theorem for UGES \cite[Theorem 4.14]{Khalil(2002)}, this enables to construct a Lyapunov function $V_e\left(\xb_e,t\right):\mathbb{R}^{d}\times\mathbb{R}_+\rightarrow\mathbb{R}$ such that 
\begin{equation}
\ba\label{eq:AC.b.Ve}
c_1\|\xb_e\|^2 \leq V_{e}\left(\xb_e,t\right)&\leq c_2\|\xb_e\|^2,\\
\frac{\partial V_{e}}{\partial t} - \frac{\partial V_{e}}{\partial \xb_e} \mathbf{C}\left(\xb_e,t\right)  &\leq -c_3\|\xb_e\|^2,\\
\left\|\frac{\partial V_{e}}{\partial \xb_e}\right\| &\leq c_4 \|\xb_e\|,
\ea
\end{equation}
for some \emph{explicit} parameters $c_1,c_2,c_3,c_4>0$.}

In the following, we show that various existing compressors can be categorized into ST or SST compressors. 

\begin{example}
    {\bf The  scalarization compressor} $\mathbf{C}_1:\mathbb{R}^d\times\mathbb{R}_+\rightarrow \mathbb{R}^d$ satisfies $\mathbf{C}_1\left(\xb_e,t\right)=\psib(t)\psib(t)^T\xb_e$, where the  compression vector $\psib:\mathbb{R}_+\rightarrow \mathbb{R}^d$ is uniformly bounded and persistently excited, 
    i.e., 
        \begin{equation}
        \label{eq:PE}
    \ba
    \alpha_2 \Ib_d \geq  \int_{t}^{t+T_1} \psib\left(s\right)\psib^T \left(s\right) ds \geq \alpha_1 \Ib_d\,,\quad \forall t\geq 0;
    \ea           
        \end{equation}
    for some constants $\alpha_1,\alpha_2,T_1>0$   (see \cite{LW-DSFN}). 
    \hfill$\square$
\end{example}


\begin{example}
{\bf The contraction compressor} $\mathbf{C}_2:\mathbb{R}^d\rightarrow \mathbb{R}^d$ satisfies 
\begin{equation}
    \label{ass_c2}
    \ba 
    \left\|\frac{\mathbf{C}_2\left(\xb_e\right)}{p}-\xb_e\right\|^2\leq \left(1-\varphi\right)\|\xb_e\|^2,
    \ea
    \end{equation} 
    for some $\varphi\in\left(0,1\right]$ and $p>0$ (see \cite{XY-CCFD,AR-AEQD,ACGT}, with the expectation operator removed\footnote{ {Readers of interest can refer to the Section V.4 for the stochastic version of $\mathbf{C}_2$.}}). The following $\mathbf{C}_{2a}$ and $\mathbf{C}_{2b}$ are specific examples of $\mathbf{C}_2$ ($p=1$ and $\varphi=\frac{k}{d}$ of $\mathbf{C}_{2a}$, $p=\frac{d}{2}$ and $\varphi=\frac{1}{d^2}$ of $\mathbf{C}_{2b}$, $p=1$ and $\varphi=\frac{3}{4}$ of $\mathbf{C}_{2c}$):
    \begin{itemize}
    \item[2a).] {\bf Greedy (Top-k) sparsifier} \cite{Beznosikov2020Onbiased}, which is given by 
$\mathbf{C}_{2a}\left(\xb_e\right)=\sum_{s=1}^{k}[\xb_e]_{i_s}{\bf e}_{i_s}$
where $i_1,\dots,i_k$ are the indices of 
largest $k$ coordinates in the absolute value of $\xb_e$.
\item[2b).] {\bf Standard uniform quantizer} \cite{XY-CCFD}, which is given by
$\mathbf{C}_{2b}\left(\xb_e\right)=\frac{\|\xb_e\|_\infty}{2}\mathrm{sgn}\left(\xb_e\right),$ 
where $\mathrm{sgn} \left(\cdot\right)$ denotes  the element-wise sign.
\item[2c).] {\bf Saturated quantizer}, which is given by 
\[[ \mathbf{C}_{2c}\left(\xb_e\right)]_i=\left\{
\ba
             [\xb_e]_i,  \quad\quad \ \ |[\xb_e]_i|\leq \Delta, \\  
             \Delta\left\lfloor\frac{[\xb_e]_i}{\Delta}\right\rfloor, \quad |[\xb_e]_i|> \Delta.\\     
             \ea  
\right. \] 
where $ i=1,2,\dots,d$, $\Delta\in\mathbb{R}$ denotes the quantization precision and $\lfloor\cdot\rfloor$ denotes the flooring function. \hfill$\square$
\end{itemize}
\end{example}

 {
Inspired by \cite{LCOC,DCWL} and the definition of ST compressor, we also propose a new compressor below.}
 {
\begin{example}
    {\bf The  scaled flooring compressor} $\mathbf{C}_3:\mathbb{R}^d\times\mathbb{R}_+\rightarrow \mathbb{R}^d$ satisfies $\mathbf{C}_3\left(\xb_e,t\right)=\gamma_e^t\left\lfloor\frac{\xb_e}{\gamma_e^t}\right\rfloor$, with $e^{-1}<\gamma_e<1$.   
    \hfill$\square$
\end{example}
}

\begin{proposition}
\label{propos}
The following statements are true:
a). $\mathbf{C}_1$ belongs to the SST compressor;
b). $\mathbf{C}_2$ belongs to the ST compressor; 
c). $\mathbf{C}_3$ belongs to the ST compressor. \hfill$\square$

\end{proposition}

\begin{remark}
\label{remark1}
 {
The new compressor $\mathbf{C}_3$ in Example 3 is established on the flooring compressor (i.e., letting $\gamma_e = 1$), by introducing an exponential scaling function $\gamma_e^t$, which enables to satisfy P1) in Definition 1. As a result, this benefits to achieve unbiased convergence (as shown in the subsequent results), in contrast to biased convergence in \cite{DSMA} where the transmitted values are also integers.  
On the other hand, it is noted from the proof of Proposition \ref{propos}. c) in Appendix \ref{app:propo} that ${\mathbf{x}_e}/{\gamma_e^t}$ is bounded for the system $\dot{\mathbf{x}}_e = -\mathbf{C}_3(\mathbf{x}_e, t)$.
}
    \hfill$\square$
\end{remark}

\begin{remark}
We stress that when the compressor $\mathbf{C}\left(\xb_e,t\right)$ is used, we do not mean to use $\mathbf{C}\left(\xb_e,t\right)$ to encode  $\xb_e$ for communication and then transmit the whole vector of $\mathbf{C}$ directly. Instead, $\mathbf{C}$ represents the communication information, whose transmission can be implemented requiring fewer bandwidths  than directly transmitting $\xb_e$ of $d$ dimensions, leading to the so-called communication compression. For example, if the scalarization compressor $\mathbf{C}_1$ is adopted, the actual communication message in each round is a \emph{scalar} $\psib(t)^T \xb_e(t)$ with each agent holding a common $\psib(t)$. For the standard uniform quantizer $\mathbf{C}_{2b}$, the actual communication message consists of a scalar $\|\xb_e\|_\infty$ and a signal vector $\mathrm{sgn}\left(\xb_e\right)$.  {For the scaled flooring compressor  $\mathbf{C}_3$, the transmitted value is an integer vector $\left\lfloor\frac{\xb_e}{\gamma_e^t}\right\rfloor$.}
In view of this, with a bit of abuse of notation, we insist on saying the mapping $\mathbf{C}$ to be a compressor throughout the paper.
\hfill$\square$
\end{remark}

\begin{remark}
In contrast with the conventional compressors, e.g., the contraction compressor, the ST compressor exhibits two distinctive features. First, it synthesizes information from both the time and space domains, broadening its applicability and expanding the design possibilities  {of compressors, such as the scalarization compressor  $\mathbf{C}_1$ and the scaled flooring compressor $\mathbf{C}_3$, both using the time information}.
Second, its key characteristic is elucidated through a non-autonomous system, which can simplify the design procedure while providing the flexibility to incorporate control-related tools into distributed optimization,  {such as the converse Lyapunov Theorem used throughout the proofs in the paper.}
 \hfill$\square$
\end{remark}

 {
{\bf Problem of Interest.} In view of the above notion of ST compressors, the following two intuitive questions can be naturally raised, which will be addressed in this paper.
\begin{itemize}
    \item[Q1).] How to incorporate ST compressors into distributed primal-dual algorithms to solve problem \eqref{eq:DO} with compressed communication?
    \item[Q2).] For the resulting distributed compressed primal-dual algorithms, can the convergence be maintained in such a way to reduce the communication burden?
\end{itemize}
}

\section{ST-Compressed Consensus Flows}
\label{sec.ave}
Distributed consensus is a fundamental algorithm that acts as a subroutine in numerous distributed optimization problems. In this section, we investigate how to combine the ST compressors with the consensus algorithm, which motivates the subsequent developments of distributed optimization algorithms with ST compressors. Moreover, due to its convenience of analysis, we focus on the continuous-time distributed consensus, taking the form
\begin{equation}
  \ba \label{eq:AC}
\dot{{\xb}}_i &= -\sum_{j\in\mathrm{N}_i} \Lb_{ij}   {\xb}_{j}.
  \ea\
\end{equation}
It is clear that  under Assumption \ref{ass-graph}, each node state exponentially reaches consensus at  ${\xb}^\ast:=\frac{1}{n}\sum_{j=0}^n \xb_j\left(0\right)$ 
\cite{CCAO}. 

\vspace{-0.2cm}
\subsection{Consensus with Direct Compression}
\label{sec-3.1}
An intuitive design of the compressed consensus algorithm
is to directly replace the information $\xb_i$ with the compressed
one for transmission, leading to the following
distributed consensus flow with direct compression (DC-DC flow) as
\begin{equation}
  \ba \label{eq:AC.a}
\dot{{\xb}}_i &= -\sum_{j\in\mathrm{N}_i} \Lb_{ij}   \mathbf{C}\left({\xb}_j,t\right).
  \ea\
\end{equation}

 {In the following, we will investigate  when the DC-DC flow \eqref{eq:AC.a} maintains exponential convergence to the average.}
Before answering this question, we make the following observation on the SST compressor. Given a SST compressor $\mathbf{C}$, it is clear that the system 
    \[
\ba 
\dot{\yb}_e=-\Lambda \overline{\Cb}\left(\yb_e,t\right),
\ea
\] 
where ${\yb}_e\in\mathbb{R}^{\left(n-1\right)d}$, $\Lambda:= \mathrm{diag}
 \left(\lambda_2,\dots,\lambda_n\right)\otimes \mathbf{I}_d
$ and $\overline{\Cb}\left(\yb,t\right):=[\mathbf{C}^T\left(\yb_{1},t\right),\dots,\mathbf{C}^T\left(\yb_{n-1},t\right)]^T$, is UGES at the zero equilibrium. By the converse Lyapunov Theorem for UGES \cite[Theorem 4.14]{Khalil(2002)}, this enables to construct a Lyapunov function $V_e:\mathbb{R}^{\left(n-1\right)d}\times\mathbb{R}_+\rightarrow\mathbb{R}_+$ such that

\begin{equation}
\begin{aligned}
    \label{eq:Ve}
\overline{c}_1\|\yb_e\|^2 \leq \overline{V}_{e}\left(\yb_e,t\right)&\leq \overline{c}_2\|\yb_e\|^2,\\
\frac{\partial \overline{V}_{e}}{\partial t} - \frac{\partial \overline{V}_{e}}{\partial \yb_e} \Lambda \overline{\Cb}\left(\yb_e,t\right)& \leq \overline{c}_3\|\yb_e\|^2,\\ 
  \left\|\frac{\partial V_{e}}{\partial \yb_e}\right\| &\leq \overline{c}_4 \|\yb_e\|,
\end{aligned}
\end{equation}
for some { explicit} constants $\overline{c}_1,\overline{c}_2,\overline{c}_3,\overline{c}_4>0$. 

With this in mind, by defining $\calSb:=\mathbf{S}\otimes\mathbf{I}_d$ and $\Cb\left(\xb,t\right):=[\mathbf{C}^T\left(\xb_{1},t\right),\dots,\mathbf{C}^T\left(\xb_{n},t\right)]^T$, we are ready to propose the following theorem for Flow \eqref{eq:AC.a}, answering the question by showing that a condition on the communication network $\mathrm{G}$ and the SST compressor $\mathbf{C}$ is still required in order to maintain exponential convergence to the average.


\begin{theorem}\label{thm-AC.a}
Let Assumption \ref{ass-graph} hold, then for the DC-DC Flow \eqref{eq:AC.a} with a { {SST compressor}} $\mathbf{C}$, if there holds
\begin{equation}
\label{propery iii}
\ba
    \|\overline{\Cb}\left(\calSb^T\xb,t\right)-\calSb^T\Cb\left(\xb,t\right)\|\leq \delta \|\calSb^T\xb\|\,,\  \forall \left(\xb,t\right)\in\mathbb{R}^{nd}\times\mathbb{R}_+
    \ea
\end{equation}
for  {$\delta<\frac{\overline{c}_3}{\overline{c}_4\lambda_n}$}, then there holds
$
\|\xb_i(t) - {\xb}^\ast\|^2 = \mathcal{O}\left(e^{-\gamma t}\right)\,
$
for $\gamma=\frac{\overline{c}_3-\overline{c}_4\delta\lambda_n}{\overline{c}_1}$.
 \hfill$\square$

\end{theorem}

From the extra condition \eqref{propery iii}, it can be seen that the SST compressor $\mathbf{C}$  {needs to} satisfy some conditions  {relying on} the network topology (see $\calSb$)  {to} ensure the exponential convergence property of the DC-DC Flow \eqref{eq:AC.a} in general. 
Notably, by taking a linear form of SST compressor $\mathbf{C}\left(\xb_e,t\right)=M(t)\xb_e$, e.g. the scalarization compressor $\mathbf{C}_1$, we note that the extra condition \eqref{propery iii} reduces to
\[
    \|[\left(\Ib_{n-1}\otimes M(t)\right)\calSb^T-\calSb^T \left(\Ib_{n-1}\otimes M(t)\right)]\xb\|\leq \delta \|\calSb^T\xb\|\,,
\]
 {which} holds for all $\left(\xb,t\right)\in\mathbb{R}^{nd}\times\mathbb{R}_+$, since $\left(\Ib_{n-1}\otimes M(t)\right)\calSb^T-\calSb^T \left(\Ib_{n-1}\otimes M(t)\right) = 0$. This immediately implies that the linear SST compressor, e.g., the scalarization compressor $\mathbf{C}_1$, is applicable to the DC-DC Flow \eqref{eq:AC.a} with no need of any extra condition, as shown in \cite{LW-DSFN}.  {On the other hand, due to the involvement of the network topology and dependence on $(\xb,t)$, the verification of \eqref{propery iii} is generally difficult for nonlinear compressors. This thus motivates the subsequent development of new compression methods that allow more general ST compressors applicable.}


\vspace{-0.2cm}
\subsection{Consensus with Observer-based Compression}
\label{sec-3.2}

In the previous section, it has been shown that the SST compressor can be directly incorporated, subject to an extra condition \eqref{propery iii}. This  poses limitations on the range of feasible compressors.  {In this section, to allow more general ST compressors to be incorporated, we propose another compression method based on distributed observer. The corresponding distributed compressed consensus takes the form}
\begin{equation}
  \ba \label{eq:AC.b}
\dot{{\xb}}_i &= -\alpha\sum_{j\in\mathrm{N}_i} \Lb_{ij}   {\hat{\xb}}^i_j, \\
  \dot{\hat{\xb}}^i_j&={\xb}_{j,c} ,\quad j\in\mathrm{N}_i,\\
{\xb}_{i,c}&=\mathbf{C}\left({\xb}_i-{\hat{\xb}}^i_i,t\right),
  \ea\
\end{equation}
where $\alpha>0$ is a gain parameter, and  ${\hat{\xb}}^j_i\left(0\right)={\hat{\xb}}^{j'}_i\left(0\right),\forall 
j,j'\in \mathrm{N}_i$, $i\in\mV$.

The proposed compressed consensus Flow \eqref{eq:AC.b} is comprised of two sets of states for each agent $i$. The state $\xb_i$ denotes the estimate of the consensus solution as in \eqref{eq:AC}, while the states ${\hat{\xb}}^i_j$ are introduced to each agent $i$ to estimate its neighboring solution state $\xb_j$, $j\in\mathrm{N}_i$. To have a better view of this, let us first ignore the compressor and have ${\xb}_{i,c}={\xb}_i-{\hat{\xb}}^i_i$ in \eqref{eq:AC.b}. Then it is clear that the ${\hat{\xb}}^i_j$ acts as an observer to estimate $\xb_j$.
 {The observer-based compression method is thus established in order to realize compression and communication of the error state ${\xb}_i-{\hat{\xb}}^i_i$. Since the compression errors for the error state ${\xb}_i-{\hat{\xb}}^i_i$ are smaller compared to directly compressing the state ${\xb}_i$, the observer-based compression method allows the use of more general compressors than direct compression.
}


\begin{theorem}\label{thm-AC.b}
Let Assumption \ref{ass-graph} hold, then for the DC-OC Flow \eqref{eq:AC.b} with a {  ST compressor} $\mathbf{C}$,  there exists $\alpha^\ast=\mathrm{min}\left\{\frac{2c_3}{9\lambda_nc_4\sqrt{n}},\frac{2c_3}{3\lambda_n}\right
\}$ such that for all $\alpha\leq \alpha^\ast$, there holds
$
\|\xb_i(t) - {\xb}^\ast\|^2 = \mathcal{O}\left(e^{-\gamma t}\right)\,,
$ 
for $\gamma=\mathrm{min}\left\{\frac{
\alpha\lambda_2}{2},\frac{ c_3}{3c_1}\right\}$.\hfill$\square$

\end{theorem}

A rigorous proof of Theorem \ref{thm-AC.b} is presented in Appendix C. Intuitively, from the perspective of control systems, we stress that the corresponding system \eqref{eq:AC.b} can be regarded as an interconnection of two subsystems: $\xb_i$-subsystem and ${\hat{\xb}}^i_j$-subsystem, with $\alpha$ a low gain that is tuned  such that the supply functions of the two interconnected subsystems satisfy some small-gain conditions for closed-loop exponential stability \cite[Theorem 5.6]{Khalil(2002)}. {  On the other hand, we note that  such $\alpha$ is not necessarily to be small, as the $\mathbf{C}$ is designable and can be chosen so as to have a large margin for $\alpha$.}

\color{black}
\vspace{-0.2cm}
\section{ST-Compressed Primal-Dual Flows}
\label{sec.CPD}

In the previous section, two compressor incorporation methods have been introduced to the consensus flow with exponential convergence guarantees. In the following, we will show that the resulting two ST-compressed consensus flows can be further explored, respectively, to establish distributed ST-compressed primal-dual flows based on Flow \eqref{eq:Primal_Dual} for problem \eqref{eq:DO} with linear convergence guarantees.

\vspace{-0.2cm}
\subsection{Direct Compression}

 In this subsection, we aim to propose a distributed compressed primal-dual flow for problem \eqref{eq:DO} based on the directly compressed consensus Flow \eqref{eq:AC.a}.
The proposed distributed primal-dual flow with direct compression takes the form
\begin{equation}
\ba \label{eq:CPD.a}
\dot{\xb}_i&=-\sum^{n}_{j=1}\Lb_{ij} \mathbf{C}\left(\xb_j,t\right)-\beta \vb_i-\eta \nabla f_i\left(\xb_i\right), \\
\dot{\vb}_i&={\beta\sum^{n}_{j=1}\Lb_{ij}\mathbf{C}\left(\xb_j,t\right)},
\ea
\end{equation}
where the initial condition  $\sum_{i=1}^n \vb_i\left(0\right)=\mathbf{0}_d$.

We propose the following theorem for Flow \eqref{eq:CPD.a}. 

\begin{theorem}\label{thm-CPD.a}
Let Assumptions \ref{ass-DO} and \ref{ass-graph} hold, and $\mathbf{C}$ be a {SST compressor} satisfying \eqref{propery iii} with  $\delta>0$. Then there exist  $\beta,\eta>0$ such that $\xb_i(t)$ generated by Flow \eqref{eq:CPD.a} converges to the optimal solution $s^\ast$ exponentially,  { i.e., $\|\xb_i(t)-s^\ast\|^2=\mathcal{O}\left(e^{-\gamma t}\right)$} (see Appendix \ref{app:thm-CPD.a} for   {explicit expressions of} parameters $\delta,\beta,\eta$ and the convergence rate $\gamma$).\hfill$\square$

\end{theorem}

This theorem demonstrates the effectiveness of direct compression. When the conditions in Theorem \ref{thm-CPD.a} are satisfied, direct compression of the compressor can ensure exponential convergence, without introducing extra states required by other compression methods. { The convergence rates presented in Theorem \ref{thm-CPD.a} and the subsequent theorems are in fact explicitly derived. However, because of the involvement of numerous intermediate variables, we provide the detailed expressions in Appendices for readers of interest.}
\subsection{Observer-based Compression}

In this section, we propose distributed compressed primal-dual  flow based on distributed observer-based compressed consensus \eqref{eq:AC.b} in Section \ref{sec-3.2}.
 
The proposed distributed primal-dual flow in continuous-time form with observer-based compression takes the form
\begin{equation}
\ba \label{eq:CPD.b}
\dot{\xb}_i&=-\alpha\sum^{n}_{j=1}\Lb_{ij} {\hat{\xb}}^i_j-\beta \vb_i-\eta \nabla f_i\left(\xb_i\right), \\
\dot{\vb}_i&={\beta\sum^{n}_{j=1}\Lb_{ij}{\hat{\xb}}^i_j},\\
  \dot{\hat{\xb}}^i_j&= {\xb}_{j,c} ,\quad j\in\mathrm{N}_i,\\
{\xb}_{i,c}&=\mathbf{C}\left({\xb}_i-{\hat{\xb}}^i_i,t\right),
\ea
\end{equation}
where the initial condition is $\sum_{i=1}^n \vb_i\left(0\right)=\mathbf{0}_d$ and for each
$i\in\mathrm{V}$, ${\hat{\xb}}^j_i\left(0\right)={\hat{\xb}}^{j'}_i\left(0\right),\forall 
j,j'\in \mV$.

We propose the following theorem for Flow \eqref{eq:CPD.b}. 

\begin{theorem}\label{thm-CPD.b}
Let Assumptions \ref{ass-DO} and \ref{ass-graph} hold, and $\mathbf{C}$ be a {ST compressor}. Then there exist $\alpha,\beta,\eta>0$ such that $\xb_i(t)$ generated by Flow \eqref{eq:CPD.b} converges to the optimal solution $s^\ast$ exponentially,  { i.e., $\|\xb_i(t)-s^\ast\|^2=\mathcal{O}\left(e^{-\gamma t}\right)$} (see Appendix \ref{app:thm-CPD.b} for the  {explicit expressions of} parameters $\alpha,\beta,\eta$ and the convergence rate $\gamma$).
\hfill$\square$

\end{theorem}

\begin{remark}
\label{rem:com}
{ 
We compare the methods of direct compression and observer-based compression in the following. The direct compression method does not require the introduction of any additional states, thus avoiding extra storage and computational burdens.
However, only SST compressors that satisfy the condition \eqref{propery iii} can be incorporated by the direct compression method.
On the other hand, the observer-based compression method allows more general ST compressors to be used while maintaining the effectiveness of the algorithm, but at the price of introducing extra states.}\hfill$\square$
\end{remark}




\section{Discrete Implementations}
\label{sec.DPD}

\subsection{ST Compressors in Discrete Time}

Based on the ST compressors in Definition \ref{def-SST}, we propose the following (strong) ST compressors in discrete time.

 {
\begin{definition}[ST compressor in discrete time]
\label{def-DSST}
The mapping $\mathbf{C}:\mathbb{R}^d\times\mathbb{N}\rightarrow \mathbb{R}^d$ is said to be a \emph{spatio-temporal (ST) compressor in discrete time}, if the following two properties hold.
\begin{itemize}
    \item[P1$^\prime$).] There exists a stepsize $\kappa_0>0$ such that the induced discrete-time non-autonomous system $\xb_e\left(t+1\right)=\xb_e(t)-\kappa_0\mathbf{C}\left(\xb_e(t),t\right)$ is uniformly globally linearly stable (UGLS) at the origin. 
    \item[P2$^\prime$).] There exists a $L_c>0$ such that 
   \begin{equation}\label{eq:ugl2}
\left\|\mathbf{C}\left(\xb_e,t\right)-\mathbf{C}\left(\xb_e',t\right)\right\| \leq L_c\left\|\xb_e-\xb_e'\right\|
    \end{equation} 
    holds
    for all $\xb_e\in\mathbb{R}^d,\xb_e'=0$ and any $t\in\mathbb{N}$. 
\end{itemize} 
Such mapping $\mathbf{C}$ is said to be a \emph{strong spatio-temporal (SST) compressor in discrete time}, if the UGLS property in P1$^\prime$) holds for all $\kappa_0\in \left(0,\kappa_0^\ast\right)$ with some $\kappa_0^\ast>0$, and  \eqref{eq:ugl2} in P2$^\prime$) holds for  all $\xb_e,\xb_e'\in\mathbb{R}^d$ and any $t\in\mathbb{N}$.
\hfill$\square$
\end{definition}}



In discrete-time cases, it should be noticed that the condition \eqref{eq:PE} of the scalarization compressor becomes 
$
    \alpha_2 \Ib_d \geq  \sum_{s=t}^{t+T_1-1} \psib\left(s\right)\psib^T \left(s\right) \geq \alpha_1 \Ib_d\,,\ \forall t\geq 0\,.
$
A specific example of discrete-time cases of $\mathbf{C}_1$, denoted by $\mathbf{C}_{1a}$, can be derived by letting $\psib(t)= \mathbf{e}_i $ with $i=1+\left(t\   \mathrm{mod}\ d\right)$ for $t\in\mathbb{N}$. 
\begin{proposition}
\label{pro-dis}
The following statements are true:
a). $\mathbf{C}_1$ belongs to the SST compressor in discrete time;
b). $\mathbf{C}_2$ belongs to the ST compressor  in discrete time;
 {
c). $\mathbf{C}_3$ belongs to the ST compressor  in discrete time.}
\hfill$\square$

\end{proposition}

\vspace{-0.2cm}
\subsection{Discretization of Compressed Primal-Dual Flows}

In practice, algorithms are always implemented in a discrete-time form. In the following, we discretize the Flow \eqref{eq:CPD.a} based on the Euler method, resulting in Algorithm \ref{DPD-DC}.

\begin{algorithm}
\caption{Distributed Primal-Dual algorithm with Direct Compression (DPD-DC)}
\label{DPD-DC}
\begin{algorithmic}

\State {\bf Initialization}: $ \kappa,\kappa_0,\beta,\eta>0$, $\vb_i\left(0\right)=\mathbf{0}_d$, $i\in\mV$.
\For{$t \in \mathbb{N}$, each node $i$}
     \State 
$\xb_{i}\left(t+1\right)=\xb_{i}(t)-\kappa_0\sum^{n}_{j=1}\Lb_{ij}\mathbf C\left(\xb_{j}(t),t\right)-\kappa\big(\beta \vb_i(t)
\newline~~~~~~~~~~~~~~~~~~~
+\eta \nabla f_i\left(\xb_i(t)\right)\big),$
\State $\vb_{i}\left(t+1\right)=\vb_{i}(t)+\kappa_0{\beta\sum^{n}_{j=1}\Lb_{ij}\mathbf C\left(\xb_{j}(t),t\right)}.$
\EndFor
\end{algorithmic}
\end{algorithm}



\begin{theorem}\label{thm-DPD.a}
Let Assumptions \ref{ass-DO} and \ref{ass-graph} hold, and $\mathbf{C}$ be a {  SST compressor} in discrete time, which  satisfies \eqref{propery iii} with some $\delta>0$. Then there exist some $ \kappa,\kappa_0,\beta,\eta>0$ such that  $\xb_i(t)$ generated by Algorithm \ref{DPD-DC} converges to the optimal solution $s^\ast$ linearly,  { i.e., $\|\xb_i(t)-s^\ast\|^2=\mathcal{O}\left((1-\gamma)^t\right)$} (see Appendix \ref{app:thm-DPD.a} for the  {explicit expressions of} parameters $ \delta,\kappa,\kappa_0,\beta,\eta$ and the convergence rate $\gamma$). 
\hfill$\square$

\end{theorem}

Next, we discretize Flow \eqref{eq:CPD.b} based on the  Euler approximation method,
yielding Algorithm \ref{DPD-OC}.

\begin{algorithm}
\caption{Distributed Primal-Dual algorithm with Observer-based Compression (DPD-OC)}
\label{DPD-OC}
\begin{algorithmic}
\State {\bf Initialization}: $ \kappa,\kappa_0,\beta,\eta>0$, $\vb_i\left(0\right)=\mathbf{0}_d$, $\xb_i\left(0\right)\in\mathbb{R}^d$, and $\xb_j^i\left(0\right)=\mathbf{0}_d$, $j\in\mathrm{N}_i$, $i\in\mV$.

\For{$t \in \mathbb{N}$, each node $i$}
    \For{each observer of node $j$}
    \State ${\hat{\xb}}^i_j\left(t+1\right) = {\hat{\xb}}^i_j(t)+\kappa_0 \xb_{j,c}(t)$.
    \EndFor
    \State Update:
     \State $\xb_{i}\left(t+1\right)=\xb_{i}(t)-\kappa\big(\sum_{j=1}^n\Lb_{ij}{\hat{\xb}}^i_j(t)+\beta \vb_{i}(t)\newline~~~~~~~~~~~~~~~~~~~+\eta \nabla f_i\left(\xb_{i}(t)\right)\big),$
 \State $\vb_{i}\left(t+1\right)=\vb_{i}(t)+\kappa{\beta\sum_{j=1}^n\Lb_{ij}{\hat{\xb}}^i_j(t)}.$
\EndFor
\end{algorithmic}
\end{algorithm}


\begin{theorem}\label{thm-DPD.b}
Let Assumptions \ref{ass-DO} and \ref{ass-graph} hold, and $\mathbf{C}$ be a {  ST compressor} in discrete time with  $\kappa_0>0$. Then there exist  $\kappa, \beta,\eta>0$ such that  $\xb_i(t)$ generated by Algorithm \ref{DPD-OC} converges to the optimal solution $s^\ast$ linearly,   { i.e., $\|\xb_i(t)-s^\ast\|^2=\mathcal{O}\left((1-\gamma)^t\right)$} (see Appendix \ref{app:thm-DPD.b} for the  {explicit expressions of} parameters $\kappa,\beta,\eta$ and the convergence rate $\gamma$). 
\hfill$\square$

\end{theorem}


\begin{remark}
\label{re:rate}    
{ 
In terms of the convergence rates of DPD-DC and DPD-OC established in Theorem \ref{thm-DPD.a} and Theorem \ref{thm-DPD.b}, it is difficult to have a rigorous comparison of which is faster, due to the complexity of the upper bound expressions of the stepsize parameters $\beta,\eta,\kappa$. In the following, a rough comparison is made with the ST compressor as $\mathbf{C}_{1a}$ for convenience. According to  Appendices \ref{app:thm-DPD.a} and \ref{app:thm-DPD.b}, the linear convergence rates of DPD-DC and DPD-OC can be, respectively, derived as
$\gamma_{DC}= \frac{1}{2} \kappa\mathrm{min}\left\{\frac{c_3\lambda_2}{4c_1\lambda_n},\frac{c_3\lambda_2}{4c_1},\beta^2,\eta\frac{\mu}{2n}\right\}$ and $\gamma_{OC}= \frac{1}{2}\kappa\mathrm{min}\left\{\frac{\lambda_2}{2}, {\beta^2}, \eta \frac{\mu}{2n},\frac{c_3}{2c_1}\right\}
$. 
Then, when $\beta,\eta$ are small, they may dominate the convergence rates, resulting in a similar convergence rate for both algorithms.
When parameters $\beta,\eta$ are relatively large, we may have ${\gamma}_{OC}> {\gamma}_{DC}$ as $\frac{\lambda_2}{2\lambda_n}<1$ and $\frac{c_3}{2c_1}<1$ by \eqref{eq:c3c1} in Appendix \ref{app:thm-DPD.b}. Thus, DPD-OC may be beneficial in terms of a faster convergence rate than DPD-DC under the ST compressor $\mathbf{C}_{1a}$, but at the price of introducing extra computation states and burden, as in Remark \ref{rem:com}.}
\end{remark}

\subsection{Comparison with Filter-based Compression}

\color{black}
The distributed primal-dual flow with filter-based compression (DPD-FC) takes the form \eqref{eq:fc}. Similar ideas can be seen in \cite{XY-CCFD,ACGT,CGTA,liu2021linear}. 

\begin{equation}
    \label{eq:fc}
    \ba 
{\sigmab}_{i}\left(t+1\right)&=\sigmab_i(t)+\kappa_0\qb_i(t), \\
{\zb}_{i}\left(t+1\right)&=\zb_i(t)+\kappa_0\big(\qb_i(t)-\sum^{n}_{j=1}\Lb_{ij}\qb_j(t)\big),\\
{\xb}_{i}\left(t+1\right)&=\xb_i(t)-\kappa\big(\sigmab_i(t)-\zb_i(t)+\sum^{n}_{j=1}\Lb_{ij}{\qb}_j(t)\\
&\quad +\beta \vb_{i}(t)+\eta \nabla f_i(\xb_i(t))\big), \\
{\vb}_{i}\left(t+1\right)&=\vb_i(t)+\kappa{\beta\big(\sigmab_i(t)-\zb_i(t)+\sum^{n}_{j=1}\Lb_{ij}{\qb}_j(t)\big)},\\
\qb_i(t) &= \mathbf{C}\left(\xb_i(t)-\sigmab_i(t),t\right).
\ea
\end{equation}

 {
The DPD-FC \eqref{eq:fc} introduces a distributed filter and a distributed integrator. The filter $\sigmab_i$ is used to track the state $\xb_i$, while the integrator $\zb_i$ tracks the term
$
\sigmab_i - \sum_{j=1}^{n}\Lb_{ij} \sigmab_j.
$
In contrast, the DPD-OC in Algorithm \ref{DPD-OC} introduces distributed observers to track the states of \emph{neighboring} nodes. As a result, both algorithms share a similar compression idea in the sense of compressing and transmitting error states.
}

For DPD-FC \eqref{eq:fc}, the ST compressors can be incorporated, leading to unbiased linear convergence. The corresponding analysis is referred to in the conference version \cite{ZR-ST}, but is omitted here due to space limitations.

\subsection{Stochastic ST Compressors and Algorithms}

It should be noticed that many literature on compressor assumption take into account the presence of randomness. Therefore, we extend the ST compressor to randomness and study its effectiveness in applications. In this section, we study the randomization of the ST compressor and the application of DPD-OC as a example.

Introduce randomness to ST compressors, we obtain the definition of Stochastic Spatio-Temporal (StST) Compressor, with focus on discrete time.

\begin{definition}[StST Compressor]\label{def-SC0C}
    Given a linearly {\bf mean-square} bounded mapping $\mathbf{C}:\mathbb{R}^d\times\mathbb{R}_+\rightarrow \mathbb{R}^d$, i.e., there exists a $L_c>0$ such that $\Eb\|\mathbf{C}(\xb_e,t)\|^2 \leq L^2_c\|\xb_e\|^2$ for all $\xb_e\in\mathbb{R}^d$ and any $t\in\mathbb{R}_+$. Then, $\mathbf{C}$ is said to be a \emph{StST compressor}, if the induced non-autonomous system $\xb_e(t+1)=\xb_e(t)-\kappa_0\mathbf{C}(\xb_e,t)$ is  uniformly globally exponentially stable at the origin in the {\bf mean-square} sense, for some stepsize $\kappa_0>0$.
\hfill$\square$
\end{definition}

The ST compressor is a special case of the StST compressor. Moreover, some compressor assumptions in literature belongs to the StST compressor.

\begin{example}
    {\bf The stochastic contractive compressor} $\mathbf{C}_3:\mathbb{R}^d\rightarrow \mathbb{R}^d$ satisfies 
\begin{equation}
    \label{ass_c2S}
    \ba
    \Eb\|\frac{\mathbf{C}_3(\xb_e)}{p}-\xb_e\|^2\leq (1-\varphi)\|\xb_e\|^2
    \ea
    \end{equation} 
    for some $\varphi\in(0,1]$ and $p>0$. By \cite{XY-CCFD}, the followings are specific examples of $\mathbf{C}_3$:
    \begin{itemize}
    \item[3a).] Unbiased $l$-bits quantizer \cite{liu2021linear} 
$$\mathbf{C}_{3a}(\xb_e)=\frac{\|\xb_e\|_{\infty}}{2^{l-1}}{\rm sign}(\xb_e)*\lfloor\frac{2^{l-1}|\xb_e|}{\|\xb_e\|_{\infty}}+\overline{\omega}\rfloor,$$
where $\overline{\omega}$ is a random perturbation
vector uniformly sampled from $[0,1]^d$.
\end{itemize}

\end{example}

\begin{proposition}
Compressor $\mathbf{C}_3$ belongs to the StST compressor.\hfill$\square$
\end{proposition}

The proof of Proposition 2 is similar to that of Proposition b). and is omitted for simplicity.

We apply the StST compressor to DPD-OC and propose the following theorem for DPD-OC. 

\begin{theorem}\label{thm-DPD.d}
Let Assumption \ref{ass-DO} and \ref{ass-graph} hold, and $\mathbf{C}$ be a {\bf StST compressor} with some $\kappa_0>0$. Then for $\kappa,\beta,\eta>0$, the mean square of $\xb_{i}(t)$ in the DPD-OC converges to the optimal solution $s^\ast$ linearly.\hfill$\square$

\end{theorem}

\section{Numerical Simulations}
\label{sec.num}

\subsection{Verification of  ST Compressors}

In this section, we verify that the compressors mentioned in this paper  {$\mathbf{C}_{1a}$, $\mathbf{C}_{2a}$ ($k=2$), $\mathbf{C}_{2b}$, $\mathbf{C}_{2c}$ ($\Delta=1$), $\mathbf{C}_{3a}$ ($\gamma_e=0.9$),} 
\color{black}
satisfy that the induced system $\dot{\xb}_e=-\mathbf{C}\left(\xb_e,t\right)$ is UGES at the zero equilibrium, thus belong to the ST compressors.

The plots in figures respectively demonstrate the exponential convergence system $\dot{\xb}_e=-\mathbf{C}\left(\xb_e,t\right)$ with different compressors, validating our conclusions in Proposition \ref{propos}.

\subsection{Simulations under Different Compression Methods}

In this section, we consider a network of $n=10$ nodes over a
circle communication graph and the dimension of the local state is $d=5$, where each edge is assigned with the same unit weight and each node holds a local function $f_i\left(\xb_i\right)=\frac{1}{2}\|\Hb^T_i \xb_i-b_i\|^2 $ with randomly generated $\mathbf{H}_i\in\mathbb{R}^d$ and $b_i\in\mathbb{R}$. Moreover, the functions $f_i\left(\xb_i\right)$ satisfy Assumption \ref{ass-DO} with $\mu>0$ and a unique optimal solution $s^\ast$. Next, we will incorporate different compression methods into algorithms and compare their effects.

We use the scalarization compressor $\mathbf{C}_{1a}$ and the greedy sparsifier compressor $\mathbf{C}_{2a}$ as examples.
In this application, we integrate DPD-DC, DPD-OC, DPD-FC,  {DPD-Choco} with $\mathbf{C}_{1a}$, and integrate DPD-OC, DPD-FC,  { DPD-Choco}  with $\mathbf{C}_{2a}$,  {where DPD-Choco is an algorithm incorporating the compression method from \cite{ALCA} into \eqref{eq:Primal_Dual}}. The plots in figures illustrate the evolution of the sum of squared distances from the current $\xb_i(t)$ to $s^\ast$, denoted as $\sum_{i=1}^n\|\xb_i(t)-s^\ast\|^2$  {with respect to iterations and transmitted bytes in each node}, respectively. It can be seen that the algorithms exhibit linear convergence to the optimal solution, verifying Theorem \ref{thm-DPD.a} and Theorem \ref{thm-DPD.b}.  {In addition, as we set the stepsize parameters large enough, the figures show that DPD-OC converges faster than DPD-DC, thereby requiring fewer transmitted bytes to achieve the same accuracy, which verifies Remark \ref{re:rate}.}

\vspace{-0.2cm}
\subsection{Simulations with Different Compressors}

{ 
Next, we investigate the performance of different specific compressors. For the above-mentioned problem, we incorporate compressors $\mathbf{C}_{1a}$,  $\mathbf{C}_{2a}$, $\mathbf{C}_{2b}$, $\mathbf{C}_{3a}$ into the DPD-OC proposed in this paper, while keeping all other parameters unchanged.  The plots in figures illustrate the evolution of $\sum_{i=1}^n\|\xb_i(t)-s^\ast\|^2$ with respect to iterations, which show the linear convergence of DPD-OC with any ST compressor, verifying Theorem \ref{thm-DPD.b}. Moreover, the number of bytes required for each iteration, the number of iterations required to achieve an accuracy of $10^{-4}$ and the total number of transmitted bytes are shown in Table \ref{tab:1}, from which we can observe that all compressors significantly reduce the total transmitted bytes. 
}

\begin{table}[http]
\caption{{ {Total transmitted bytes to reach $10^{-4}$ accuracy in DPD-OC with different compressors.}}}
    \centering
    \begin{tabular}{c||c|c|c|c|c}
    Compressors & No $\mathbf{C}$   & $\mathbf{C}_{1a}$ & $\mathbf{C}_{2a}$& $\mathbf{C}_{2b}$&$\mathbf{C}_{3a}$ \\ \hline Bytes for each iteration& 40&8&16&9&20\\ \hline Number of iterations /$10^4$ & 0.45 & 1.15 & 0.60 & 1.34&  0.71\\ \hline Total bytes /$10^4$& 18.0&9.2 &9.6&12.0&14.2
    \end{tabular}
     \vspace*{-0.5cm}
    \label{tab:1}
\end{table}

\subsection{Simulations with  Convex Objective Functions}

{\color{black}
Next, we discuss the case where the objective function is convex but not strongly convex, while other settings are the same as that in the previous subsection. We take the objective function from \cite{Standard_function} as}
\[
\ba 
\min\sum_{i=1}^nf_i\left(\xb\right) = \sum_{j=1}^{d-1} \left[ 100\left([\xb]_{j+1} - [\xb]_j^2\right)^2 + \left([\xb]_j - a_i\mathbf{1}_d\right)^2 \right],
\ea
\]
whose optimal solution is  $s^\ast=\mathbf{1}_d$.
The simulation results of DPD-DC, DPD-OC, DPD-FC,  {DPD-Choco}   with $\mathbf{C}_{1a}$, as examples, are shown in figures. From the figure, we can see that the above algorithm can achieve asymptotic convergence to the optimal solution for convex but not strongly convex functions.  {Moreover, in this case, there is no significant difference in the convergence rates of the different compression methods. This contrasts with the results in figures and is worth studying in our future work.}

\section{Conclusion}
\label{sec.con}

\color{black}
In this paper, we have introduced a type of spatio-temporal compressor that integrates both spatial and temporal characteristics, and effectively compresses information by leveraging information from both the time and space domains. This type of compressor has covered several compressors in the literature { and inspired the proposal of new compressors}. Our proposed compressor has been implemented in the primal-dual algorithm by the direct compression method and the observer-based compression method {  for exponential/linear convergence.  
Future work includes the application of the ST compressor to other classical distributed optimization algorithms, such as those based on stochastic gradient methods \cite{SGD}, to further explore and verify its general applicability. In addition, methods beyond compressors, such as stochastic communication \cite{StoCom} and event-triggered communication \cite{evtr}, can be integrated with our ST compressors to reduce the communication burden. Furthermore, extending the proposed algorithms to nonconvex optimization problems, particularly those with objective functions satisfying the P-Ł condition \cite{PL}, presents another promising avenue for future research.}

\appendices

\section{Proof of Proposition \ref{propos}} 
\label{app:propo}

\color{black}
\emph{Proof of a).} First, we prove $\mathbf{C}_1$ belong to ST compressors. The proof of P1) is obvious that the system $\dot{\xb}_e=-k\psib(t)\psib(t)^T\xb_e$ is UGES at the zero equilibrium for any $k>0$ by recalling \cite{Brian-TAC-1977}, and the proof of P2) can be shown by noting that $\psib(t)$ is uniformly bounded.

\emph{Proof of b).} Next, for $\mathbf{C}_2$, we
note that the property \eqref{ass_c2} is equivalent to
\begin{equation}
\ba 
\label{ass_c2'} \|{\mathbf{C}_2\left(\xb_e\right)}/{p}\|^2-2{\xb_e^T\mathbf{C}_2\left(\xb_e\right)}/{p}\leq -\varphi\|\xb_e\|^2\,.
\ea
\end{equation}
First, we prove that the system 
    $\dot{\xb}_e=-\mathbf{C}_2\left(\xb_e,t\right)$ is UGES at the zero equilibrium. By choosing the Lyapunov function $V_e\left(\xb_e\right)={\|\xb_e\|^2}/{p}$ and using \eqref{ass_c2'}, we have
$
\dot{V}_e=-2\frac{\xb_e^T\mathbf{C}_2\left(\xb_e\right)}{p}\leq -\varphi\|\xb_e\|^2.
$
With $\varphi>0$, we conclude that $\xb_e$-system is UGES at the zero equilibrium and then P1) is proved.
In addition, by \eqref{ass_c2'} and  the Young's inequality, we have
\begin{equation}
    \label{eq:c2L}
    \ba 
\|{\mathbf{C}_2\left(\xb_e\right)}/{p}\|^2&\leq  \frac{1}{2}\|{\mathbf{C}_2\left(\xb_e\right)}/{p}\|^2-\left(\varphi-2\right)\|\xb_e\|^2\\
\Rightarrow
\|{\mathbf{C}_2\left(\xb_e\right)}\|&\leq  p\sqrt{{2\left(2-\varphi\right)}}\|\xb_e\|\leq 2p\|\xb_e\|,
    \ea
\end{equation}
where the last inequality is obtained by $\varphi\in(0,1]$. Thus P2) is proved with $L_c=2p>0$.

\emph{Proof of c).}
Finally, for $\mathbf{C}_3$, to prove the system $\dot{\xb}_e=-\mathbf{C}_3(\xb_e,t)=-\gamma_e^t\left\lfloor\frac{\xb_e}{\gamma_e^t}\right\rfloor$ is UGES, we let $\zb_e(t):=\frac{\xb_e(t)}{\gamma_e^t}$, and then have
\[
\ba
\dot{\zb}_e=-\left\lfloor\zb_e\right\rfloor-\ln(\gamma_e)\zb_e.
\ea
\]
By choosing  $V_e\left(\zb_e\right)={\|\zb_e\|^2}/{2}$, we have
\[
\ba
\dot{V}_e&=-\zb_e^T\left\lfloor\zb_e\right\rfloor-\ln(\gamma_e)\zb_e^T\zb_e\\&\leq -(1+\ln(\gamma_e))\|\zb_e\|^2+\sqrt{d}\|\zb_e\|\\
&\leq-(1+\ln(\gamma_e))V_e+\frac{d}{2(1+\ln(\gamma_e))}.
\ea
\]
As $\gamma_e\in(e^{-1},1)$, we can obtain  $1+\ln(\gamma_e)>0$. Thus
 $V_e(t)$ and  $\zb_e(t)$ are bounded uniformly. With $\xb_e(t)=\zb_e(t)\gamma_e^t$ and $\gamma_e< 1$, we complete the proof of P1) with $k=1$. In addition, it is easy to obtain that $\|\mathbf{C}_3(\xb_e,t)\|\leq \|\xb_e\|$ for all $t\geq0$, and thus P2) is proved with $L_c=1$. The proof of Proposition \ref{propos} is completed.


\color{black}
\section{Proof of Theorem \ref{thm-AC.a}}
\label{app:thm-AC.a}
Flow \eqref{eq:AC.a} can be written in a compact form as 
\begin{equation}\label{eq:AC.a.1}
    \ba 
    \dot{\xb}&=-\mathbf{L}_\otimes\xb_c,\\
    \xb_c&= \Cb\left(\xb,t\right),
    \ea
\end{equation}
where $\xb:=[\xb_1^T,\dots,\xb_n^T]^T$,  $\xb_c:=[\xb_{1,c}^T,\dots,\xb_{n,c}^T]^T$ and $\mathbf{L}_\otimes:=\Lb\otimes\mathbf{I}_d$. 
We decompose $\xb$  by defining $\xb_\perp:=\calSb^T\xb=[\xb^T_{\perp,1},\dots,\xb^T_{\perp,n-1}]^T$ and $\xb_\parallel:=\onb^T\xb$, 
where  $\onb:=\frac{1}{\sqrt{n}}\mathbf{1}_n\otimes \mathbf{I}_d$. 
This immediately implies $\dot{\xb}_\parallel=\mathbf{0}_{d}$ using the fact
\begin{equation}
\ba\label{eq:HL}
\onb^T \mathbf{L}_\otimes=\mathbf{0}\quad \mathbf{L}_\otimes\onb=\mathbf{0}.
\ea
\end{equation}
Moreover, with 
\begin{equation}
 \ba
     \label{eq:HK}\calSb\calSb^T+\onb\onb^T=\mathbf{I}_{nd},
     \ea
 \end{equation}
it is clear that $\xb_i(t)$ converges to the average consensus exponentially if  $\xb_\perp(t)$ is shown to be convergent to zero exponentially. 

With the above in mind, we compute the time derivative of $\xb_\perp$ as 
\begin{equation}\label{eq:AC.a.2}
 \ba 
\dot{\xb}_\perp= -\calSb^T\mathbf{L}_\otimes\mathbf{\Cb}\left(\xb,t\right),
\ea   
\end{equation}
for which  we choose a Lyapunov function  as  $V\left(\xb_\perp,t\right):=\overline{V}_e\left(\xb_\perp,t\right)$, which is defined in \eqref{eq:Ve}, and obtain
\begin{equation}
\ba 
\label{eq:AC.a.V}
\dot{V}&=\frac{\partial V}{\partial t} - \frac{\partial V}{\partial \xb_\perp} \Lambda \overline{\Cb}\left({\xb}_\perp,t\right)+\frac{\partial V}{\partial \xb_\perp}[\Lambda \overline{\Cb}\left({\xb}_\perp,t\right)-\calSb^T\mathbf{L}_\otimes\mathbf{\Cb}\left(\xb,t\right)] \\
&\leq -\left(\overline{c}_3-\overline{c}_4\delta\lambda_n\right)\|\xb_\perp\|^2,
\ea
\end{equation}
where the  inequality is obtained by using
\[
\ba 
\|\Lambda \overline{\Cb}\left(\yb_e,t\right)-\calSb^T\mathbf{L}_\otimes\mathbf{\Cb}\left(\xb,t\right)\|
&\leq \lambda_n\|\overline{\Cb}\left({\xb}_\perp,t\right)-\calSb^T\Cb\left(\xb,t\right)\|\\
&\leq\delta\lambda_n\|\xb_\perp\|.
\ea
\]
For $\delta<\frac{\overline{c}_3}{\overline{c}_4\lambda_n}$, $\dot{V}$ is negative  definite. With \eqref{eq:Ve}, 
we further have
$
\dot{V}\leq -\frac{\overline{c}_3-\overline{c}_4\delta\lambda_n}{\overline{c}_1}V,
$
yielding $\|\xb_\perp(t)\|^2=\mathcal{O}\left(e^{-\gamma t}\right)$ with $\gamma=\frac{\overline{c}_3-\overline{c}_4\delta\lambda_n}{\overline{c}_1}$. The theorem is thus proved. 

\section{Proof of Theorem \ref{thm-AC.b}}
\label{app:thm-AC.b}
From Flow \eqref{eq:AC.b} and its initial condition, we can obtain that for each 
$i\in \mV$, ${\hat{\xb}}^i_j\left(0\right)={\hat{\xb}}^{j'}_i\left(0\right),\forall 
j,j'\in \mV$, i.e., the stored value of $\xb_i$ is  same in each node. Thus the stored value of each node can be written as $\xb_c:=[\xb_{1,c}^T,\dots,\xb_{n,c}^T]^T$. Then Flow \eqref{eq:AC.b} can be written in a compact form as
\begin{equation}\label{eq:AC.b.1}
    \ba 
    \dot{\xb}&=-\alpha\mathbf{L}_\otimes \xb_c,
    \\\dot{\xb}_c&=\Cb\left(\xb-\xb_c, t\right),
    \ea
\end{equation}
where $\Cb\left(\xb-\xb_c,t\right)=[\mathbf{C}^T\left(\xb_{1}-\xb_{1,c}, t\right),\dots,\mathbf{C}^T\left(\xb_{n}-\xb_{n,c}, t\right)]^T.$
Similarly, we decompose $\xb$  by defining $\xb_\perp:=\calSb^T\xb$ and $\xb_\parallel:=\onb^T\xb$. Then there holds $\dot{\xb}_\parallel=\mathbf{0}_d $, and the proof is done if we show that $\xb_\perp(t)$ exponentially converges to zero. 

By \eqref{eq:AC.b.1}, we have 
\begin{equation}\label{eq:AC.b.2}
    \ba 
    \dot{\xb}_\perp&=-\alpha\calSb^T\mathbf{L}_\otimes \xb_c,
    \\\dot{\xb}_c&=\Cb\left(\xb-\xb_c, t\right).
    \ea
\end{equation}

Next, we will introduce Lyapunov functions for system \eqref{eq:AC.b.2}.
By choosing $V_1\left(\xb_\perp\right):=\frac{1}{2}\|\xb_\perp\|^2$, there holds
\begin{equation}\label{eq:AC.b.V1}
    \ba 
    \dot{V}_1\leq \frac{\alpha}{2}\left(-\lambda_2\|\xb_\perp\|^2+\lambda_n\|\xb-\xb_c\|^2\right).
    \ea
\end{equation}

Letting $V_2\left(\xb-\xb_c,t\right):=\sum_{i=1}^n V_e\left(\xb_i-\xb_{i,c}, t\right)$, which is defined in \eqref{eq:AC.b.Ve}, then we have
\begin{equation}\label{eq:AC.b.V2}
    \ba 
    \dot{V}_2&\leq -c_3\|\xb-\xb_c\|^2+\frac{\alpha}{2}\lambda_nc_4\sqrt{n} \|\xb-\xb_c\|^2 + \frac{\alpha}{2} c_4\sqrt{n} \xb_c^T\Lb_\otimes\xb_c
    \\
    &\leq -\left(c_3-\frac{3\alpha}{2}\lambda_nc_4\sqrt{n}\right)\|\xb-\xb_c\|^2+\alpha \lambda_n c_4\sqrt{n}\|\xb_\perp\|^2,
    \ea
\end{equation}
where the first inequality is obtained by \eqref{eq:AC.b.Ve} and the second inequality is obtained by the fact
\[
\ba 
\xb_c^T\Lb_\otimes\xb_c\leq2\lambda_n\|\xb-\xb_c\|^2 + 2\lambda_n\|\xb_\perp\|^2.
\ea
\]
Define the Lyapunov function for system \eqref{eq:AC.b.2} as $V:=\chi_0V_1+V_2$ with $\chi_0=\frac{4\lambda_nc_4\sqrt{n}}{\lambda_2}$. Then for any given $\alpha\leq\alpha^\ast:=\mathrm{min}\left\{\frac{2c_3}{9\lambda_nc_4\sqrt{n}},\frac{\lambda_2c_3}{6\lambda^2_nc_4\sqrt{n}}\right\}$, and with \eqref{eq:AC.b.V1} and \eqref{eq:AC.b.V2}, we have
\begin{equation}\label{eq:AC.b.V}
\ba 
\dot{V}\leq -\frac{\alpha}{4}\chi_0\lambda_2\|\xb_\perp\|^2-\frac{c_3}{3}\|\xb-\xb_c\|^2.
\ea
\end{equation}
With \eqref{eq:AC.b.Ve}, we have
$
V\geq \frac{\chi_0}{2}\|\xb_\perp\|^2+c_1\|\xb-\xb_c\|^2,
$
then  $\|\xb_\perp(t)\|^2=\mathcal{O}\left(e^{-\gamma t}\right)$ with $\gamma=\mathrm{min}\left\{\frac{
\alpha\lambda_2}{2},\frac{ c_3}{3c_1}\right\}$. The theorem is thus proved. 

\color{black}
\section{Proof of Theorem \ref{thm-CPD.a}}
\label{app:thm-CPD.a}

Flow \eqref{eq:CPD.a} can be written in a compact form as 
\begin{equation}
\ba \label{eq:CPD.a.1}
\dot{\xb}&= -\mathbf{L}_\otimes{\xb_c}-\beta \vb-\eta \Fb\left(\xb\right), \\
\dot{\vb}&={\beta \mathbf{L}_\otimes\xb_c},\\
\xb_c&= \Cb\left(\xb,t\right),
\ea
\end{equation}
where $\vb:=[\vb_{1}^T,\dots,\vb_{n}^T]^T
$ and $\Fb\left(\xb\right):=[\nabla f_1^T\left(\xb_{1}\right),\dots,\nabla f_n^T\left(\xb_{n}\right)]^T
$.

As $f\left(x\right)$ is strongly convex, there exists a unique $s^\ast\in\mathbb{R}^d$ such that $\nabla f\left(s^\ast\right)=\mathbf{0}_d$, i.e., $\onb^T \Fb\left(\calSo\right)=\mathbf{0}_d$ with $\calSo:=\sqrt{n}\onb s^\ast$. Then it can be easily verified that  $(\xb,\vb)=(\calSo,-\frac{\eta \Fb\left(\calSo\right)}{\beta})$ is the equilibrium point of system \eqref{eq:CPD.a.1}. 
Define state errors  $\overline{\xb}:=\xb-\calSo$ and $\overline{\vb}:=\vb+\frac{\eta \Fb\left(\calSo\right)}{\beta}$, whose time derivatives along \eqref{eq:CPD.a.1} are given by
\begin{equation}
\ba \label{eq:CPD.a.2}
\dot{\xl}&=-\mathbf{L}_\otimes\xb_c-\beta \vl-\eta \Fl\left(\xl\right), \\
\dot{\vl}&=\beta \mathbf{L}_\otimes\xb_c,\\
\xb_c&=\Cb\left(\xb,t\right),
\ea
\end{equation}
where $\Fl\left(\xl\right):=\Fb\left(\xl+\calSo\right)-\Fb\left(\calSo\right)$.

We decompose $\xl$ and $\vl$ by defining $\xchc:=\calSb^T\xl$, $\xpic:=\onb^T\xl$, $\vchc:=\calSb^T\vl$ and  $\vpic:=\onb^T\vl$.
From \eqref{eq:HK}, it can be concluded that the exponential convergence of $\xl(t)$ and $\vl(t)$ is proved if $\xpic(t)$, $\xchc(t)$, $\vpic(t)$ and $\vchc(t)$ are shown to exponentially converge to zero.

Now we proceed to investigate exponential convergence of $\xpic(t)$, $\xchc(t)$, $\vpic(t)$ and $\vchc(t)$. From  \eqref{eq:CPD.a.1} and \eqref{eq:HL}, it is clear that $\onb^T\dot\vb(t)=\mathbf{0}_{d}$. With the initial condition  $\onb^T\vb\left(0\right)=\mathbf{0}_{d}$,  we have 
\begin{equation}
\label{eq:Hv}
    \ba 
\vpic(t)=\onb^T\vl(t)=\onb^T(\vb(t)-\frac{\eta\Fb\left(\calSo\right)}{\beta})=\mathbf{0}_{d}.
    \ea
\end{equation}
Then, taking the time derivatives of $\xpic(t)$, $\xchc(t)$ and $\vchc(t)$, yields
\begin{equation}
\ba \label{eq:CPD.a.3}
\mathbf{\dot{\overline{x}}}_{\perp}&=-\calSb^T \mathbf{L}_\otimes \xb_c-\beta \vchc -\eta\calSb^T \Fl\left(\xl\right),\\
\mathbf{\dot{\overline{x}}}_{\parallel}&=-\eta\onb^T\Fl\left(\xl\right),\\
\mathbf{\dot{\tilde{v}}}_{\perp}&=\beta\calSb^T \mathbf{L}_\otimes \xb_c,\\
\xb_c&=\Cb \left(\xb,t\right)-\Cb\left(\onb\xpic+\calSo,t\right),
\ea
\end{equation}
where \eqref{eq:HL}, \eqref{eq:Hv}, and  $\mathbf{L}_\otimes\Cb\left(\onb\xl+\calSo,t\right)=\mathbf{0}_{nd}$ are used.



Consider the change of coordinate $\zb:=\frac{1}{\beta}\vchc+\xchc$. System \eqref{eq:CPD.a.3} thus can be equivalently transformed into the one under coordinates $({\overline{\xb}}_{\perp}, {\overline{\xb}}_{\parallel}, {\zb})$, as
\begin{equation}
\label{eq:CPD.a.4}
\ba 
\mathbf{\dot{\overline{x}}}_{\perp}&=-\calSb^T \mathbf{L}_\otimes \xb_c+\beta^2\xchc-\beta^2 \zb -\eta\calSb^T \Fl\left(\xl\right),\\
\mathbf{\dot{\overline{x}}}_{\parallel}&=-\eta\onb^T\Fl\left(\xl\right),\\
    \dot{\zb}&=-\beta^2 \zb +\beta^2\xchc- \eta\calSb^T \Fl\left(\xl\right),\\
    \xb_c&=\Cb \left(\xb,t\right)-\Cb\left(\onb\xpic+\calSo,t\right) .
    \ea
\end{equation}
In the following, the stability of system \eqref{eq:CPD.a.4} will be investigated.
Let $V_{1}\left(\xchc,\zb\right)=\frac{1}{2}\left(\|\xchc\|^2+\|\zb\|^2\right)$, whose time derivative is given by
\begin{equation}
    \label{eq:CPD.a.V1}
\ba 
\dot{V}_{1} &\leq  
-\xchc^T\calSb\mathbf{L}_\otimes\xb_c
-\beta^2 \|\zb\|^2
+\beta^2 \|\xchc\|^2
 \\
&\quad -\eta \zb^T\calSb^T \Fl\left(\xl\right)-\eta\xch^T\calSb^T\Fl\left(\xl\right)]\\
&\leq L_c\lambda_n\|\xchc\|^2-\left(\beta^2-\frac{\eta}{2}\right) \|\zb\|^2 \\
&\quad +\left(\beta^2+\frac{\eta}{2}+{\eta}L_f^2\right) \|\xchc\|^2
+\eta L_f^2\|\xpic\|^2,
\ea
\end{equation}
where the second inequality is obtained by using
\begin{equation}
    \label{eq:Fl}
    \ba 
    \|\Fl\left(\xl\right)\|^2\leq L^2_f\left(\|\xchc\|^2+\|\xpic\|^2\right)\,,
    \ea
\end{equation}
and the fact
\begin{equation}
    \label{eq:xc}
\ba 
\|\xb_c\|&=\|\Cb \left(\xb,t\right)-\Cb\left(\onb\xpic+\calSo,t\right)\|\\
&\leq L_c\|\xb-\onb\xpic-\calSo\|
\leq L_c\|\xchc\|,
\ea  
\end{equation}
which is derived from P2) of the ST compressor $\mathbf{C}$.

Let $V_2\left(\xchc,t\right)=\overline{V}_e\left(\xchc,t\right)$ with $V_e$ defined in \eqref{eq:Ve}, whose time derivative is given by
\begin{equation}
\label{eq:CPD.a.V2}
\ba 
\dot{V}_2&=\frac{\partial V_2}{\partial t} - \frac{\partial V_2}{\partial \xchc} \Lambda \overline{\Cb}\left(\xchc,t\right)+\frac{\partial V_2}{\partial \xchc}(\Lambda \overline{\Cb}\left(\xchc,t\right)\\
&\quad -\calSb^T\mathbf{L}_\otimes\mathbf{\Cb}\left(\xb,t\right))+\frac{\partial V_2}{\partial \xb_\perp}\left(\beta^2\xchc-\beta^2 \zb -\eta\calSb^T \Fl\left(\xl\right)\right)\\
&\leq -\left(c'_3-\overline{c}_4\beta^2-\overline{c}_4\beta^2/r-\overline{c}_4\eta/r-\overline{c}_4\eta L_f^2r\right)\|\xchc\|^2\\
&\quad +\overline{c}_4\beta^2r\|\zb\|^2+\overline{c}_{4}\eta rL_f^2\|\xpic\|^2,
\ea  
\end{equation}
with $c'_3:=\overline{c}_3-{\overline{c}_4\delta\lambda_n}>0$ by choosing $\delta<\frac{c_3\lambda_2}{c_4\lambda_n}$, and $r>0$ to be determined later, where the inequality is obtained by \eqref{eq:AC.a.V}, the fact $\calSb^T\xb=\calSb^T\xl=\xchc$, \eqref{eq:Fl} and the Young's Inequality.

Let $V_{3}\left(\xpic\right)=\frac{1}{2}\|\xpic\|^2$. As $f\left(x\right)$ is $\mu$-strongly convex, we have 
\begin{equation}
    \label{eq:CPD.a.V3}
    \ba 
    \dot {V}_{3}
    &= -\eta \xpi^T\onb^T \Fl\left(\xl\right)\\
    &=-\eta\xpic^T\onb^T[\Fb\left(\xl+\calSo\right)-\Fb\left(\calSb\xchc+\calSo\right)\\
    &\quad +\Fb\left(\calSb\xchc+\calSo\right)-\Fb\left(\calSo\right)]\\&\leq -\frac{\eta\mu_n}{2} \|\xpic\|^2+\frac{\eta}{2\mu_n}L_f^2 \|\xchc\|^2,
    \ea
\end{equation}
where the  inequality is obtained by  Properties i) and ii) of $f\left(x\right)$ in Assumption 1 with $\mu_n:=\frac{\mu}{n}$.

For convenience of subsequent analysis, we introduce some parameters, which are independent of  $\beta$, $r$ and $\eta$, as 
\[
\ba 
&\chi_0=2L_c\lambda_n/c'_3,\quad
\chi_1=\frac{4L_f^2}{\mu_n}+\frac{4\chi_0\overline{c}_4L_f^2}{\mu_n},\\
&\xi_1=\frac{3}{2}+L_f^2+\chi_0\left(\overline{c}_4+\overline{c}_4L_f^2\right)+\chi_1\frac{L_f^2}{2\mu_n},\\
&\xi_2=2\chi_0\overline{c}_4,\quad
\xi_3=\chi_0\overline{c}_4,\quad
\xi_4=\frac{1}{2}.
\ea
\]

In view of the previous analysis and definitions, we choose the Lyapunov function of system \eqref{eq:CPD.a.4} as $ V = V_{1}+\chi_0V_{2}+\chi_1V_{3}$, which satisfies
\begin{equation}
\ba \label{eq:CPD.a.Vp}
V\geq \left(\frac{1}{2}+\chi_0c_1\right)\|\xchc\|^2+\frac{1}{2}\|\zb\|^2+ \frac{\chi_1}{2}\|\xpic\|^2.
\ea
\end{equation}
Combining \eqref{eq:CPD.a.V1}, \eqref{eq:CPD.a.V2},   \eqref{eq:CPD.a.V3}, and letting $r\leq1,\eta\leq\beta^2$, we can further derive
\[\ba 
    \dot{V}&\leq 
     -\left(\frac{1}{2}\chi_0c'_{3}-\xi_1\beta^2-\xi_2\beta^2/r\right)\|\xchc\|^2
     \\
    &\quad -\left(\beta^2-\xi_3\beta^2r-\xi_4\eta\right)\|\zb\|^2 -\left(\eta \frac{\mu_n}{4}\chi_1\right) \|\xpic\|^2.
    \ea
\]
Thus by fixing $r=\mathrm{min}\left\{\frac{1}{4\xi_3},1\right\}$, $\beta^2\leq \mathrm{min}\left\{\frac{\chi_0c'_{3}\lambda_2}{8\xi_1}, \frac{\chi_0c'_{3}r}{8\xi_2}\right\}$, $\eta\leq \mathrm{min}\left\{\beta^2,\frac{\beta^2}{4\xi_4}\right\}$, it can be verified that
 $\dot{V}$ is negative definite. With  \eqref{eq:CPD.a.Vp}, we have {
\[
\dot{V}\leq -\gamma V,\quad \gamma = \mathrm{min}\left\{\frac{L_c\lambda_n\lambda_2c_3'}{\lambda_2c_3'+4L_c\lambda_n\overline{c}_1},\beta^2,\eta\frac{\mu}{2n}\right\}.
\]}
Therefore, we have $\|\xl(t)\|^2=\mathcal{O}\left(e^{-\gamma t}\right)$ by recalling the definition of $V(t)$.  This immediately implies by $\xl(t):=\xb(t)-\calSo$ that $\xb_i(t)$ in Flow \eqref{eq:CPD.a} converges exponentially to the optimal solution $s^\ast$  with the SST compressor. The proof is completed.

\section{Proof of Theorem \ref{thm-CPD.b}}
\label{app:thm-CPD.b}
As analyzed in Appendix \ref{app:thm-AC.b}, Flow \eqref{eq:CPD.b} satisfies that for each 
$i\in \mV$, there holds ${\hat{\xb}}^j_i\left(0\right)={\hat{\xb}}^{j'}_i\left(0\right),\forall 
j,j'\in \mV$. Then Flow \eqref{eq:CPD.b} can be written as 
\begin{equation}\label{eq:CPD.b.1}
    \ba 
    \dot{\xb}&= -\alpha\mathbf{L}_\otimes{\xb_c}-\beta \vb-\eta \Fb\left(\xb\right), \\
\dot{\vb}&={\beta \mathbf{L}_\otimes\xb_c},\\\dot{\xb}_c&=\Cb\left(\xb-\xb_c,t\right).
    \ea
\end{equation}
We carry out a similar proof process by defining 
$\xpic(t)$, $\xchc(t)$, $\vpic(t)$ and $\vchc(t)$ described in Appendix \ref{app:thm-CPD.a}. Consider the change of coordinate   $\zb:=\frac{\alpha}{\beta}\vchc+\xchc$, and we yield the following system for $({\overline{\xb}}_{\perp}, {\overline{\xb}}_{\parallel}, {\zb})$, as
\begin{equation}
\label{eq:CPD.b.2}
\ba 
\mathbf{\dot{\tilde{x}}}_{\perp}&=-\alpha\calSb^T \mathbf{L}_\otimes \xhl+\beta^2_\alpha\xchc-\beta^2_\alpha \zb -\eta\calSb^T \Fl\left(\xl\right),\\
\mathbf{\dot{\tilde{x}}}_{\parallel}&=-\eta\onb^T\Fl\left(\xl\right),\\
    \dot{\zb}&=-\beta^2_\alpha \zb +\beta^2_\alpha\xchc- \eta\calSb^T \Fl\left(\xl\right),\\
    \dot{\tilde{\xb}}_c&=\Cb\left(\xl-\xhl,t\right), 
    \ea
\end{equation}
where $\xhl:=\xb_c-\calSo$ and $\beta^2_\alpha:=\beta^2/\alpha$.

In the following, the stability of system \eqref{eq:CPD.b.2} will be investigated.
Let $V_{1}\left(\xchc,\zb\right)=\frac{1}{2}\left(\|\xchc\|^2+\|\zb\|^2\right)$, whose time derivative is given by
\begin{equation}
    \label{eq:CPD.b.V1}
\ba 
\dot{V}_{1} &\leq  
-\alpha\xchc^T\calSb\mathbf{L}_\otimes\xhl
-\beta^2_\alpha \|\zb\|^2
+\beta^2_\alpha \|\xchc\|^2
 \\
&\quad -\eta \zb^T\calSb^T \Fl\left(\xl\right)-\eta\xch^T\calSb^T\Fl\left(\xl\right)]\\
&\leq -\frac{\alpha\lambda_2}{2}\|\xchc\|^2-\left(\beta^2_\alpha-\frac{\eta}{2}\right) \|\zb\|^2 \\
&\quad +\left(\beta^2_\alpha+\frac{\eta}{2}+{\eta}L_f^2\right) \|\xchc\|^2
+\eta L_f^2\|\xpic\|^2\\
&\quad +\frac{\alpha\lambda_n}{2}\|\xl-\xhl\|^2,
\ea
\end{equation}
where the second inequality is obtained by the fact
\begin{equation}
    \label{eq:xSLx}
    \ba 
    -\xchc^T\calSb^T\mathbf{L}_\otimes \xhl&\leq -\frac{1}{2}\lambda_2\left(\|\xchc\|^2\right)+\frac{1}{2}\lambda_n\left(\|\xl-\xhl\|^2\right).
    \ea
\end{equation}
For $V_2\left(\xl-\xhl,t\right)=V_2\left(\xb-\xb_c,t\right):=\sum_{i=1}^n V_e\left(\xb_i-\xb_{i,c}, t\right)$, which is defined in \eqref{eq:AC.b.Ve}, we have
\begin{equation}
    \label{eq:CPD.b.V2}
    \ba 
    \dot{V}_{2}
    &\leq-c_3\|\xl-\xhl\|^2 + c_4\sqrt{n}\|\xl-\xhl\|\|\alpha \mathbf{L}_\otimes\xhl\\
    &\quad +\beta^2_\alpha \calSb\zb_{k}-\beta_\alpha^2\calSb\xchc+\eta \Fl\left(\xl\right)\|
    \\
    &\leq-\left[ c_3- c_4\sqrt{n}\left( \frac{ \alpha}{r}+\frac{2
 \beta_\alpha^2}{r}+\frac{\eta}{r}+2 \alpha r\lambda_n^2 \right)\right]\|\xl-\xhl\|^2 \\
&\quad  +c_4\sqrt{n} \beta^2_\alpha r\|\zb\|^2+\left(2c_4\sqrt{n} \alpha r\lambda_n^2+c_4\beta^2_\alpha r\right)\|\xchc\|^2\\
&\quad +c_4\sqrt{n} \eta rL_f^2\|\xchc\|^2+c_4\sqrt{n} \eta rL_f^2\|\xpic\|^2,
    \ea
\end{equation}
where $r>0$ is a parameter to be determined later, the first inequality is obtained by $\vl=\calSb\vchc$ and \eqref{eq:AC.b.Ve}, and the last inequality is obtained by \eqref{eq:Fl}, the fact 
\begin{equation}
    \label{eq:CPD.b.fact}
\ba 
\xb_c^T{\Lb^2_\otimes}\xb_c&\leq2\lambda_n^2\|\xl-\xhl\|^2 + 2\lambda_n^2\|\xchc\|^2,
\ea
\end{equation}
and the Young's Inequality.

For convenience of subsequent analysis, let us introduce some positive parameters, independent of  $\alpha$, $\beta$, $r$ and $\eta$, as
\[
\ba 
&\quad \chi_1=\frac{4L_f^2}{\mu_n}+\frac{4c_4\sqrt{n}L_f^2}{\mu_n},\quad\xi_1=\frac{\lambda_2}{2},\\
&\quad \xi_2=\frac{3}{2}+L_f^2+c_4\sqrt{n}+c_4\sqrt{n}L_f^2+\chi_1\frac{L_f^2}{2\mu_n},\\
&\quad \xi_3=2c_4\sqrt{n} \lambda_n^2,\quad
\xi_4=\frac{1}{2},\quad
\xi_5=c_4\sqrt{n},\\&\quad 
\xi_6=4c_4\sqrt{n},\quad
\xi_7=\frac{\lambda_n}{2}+2c_4\sqrt{n}\lambda_n^2.
\ea
\]

In view of the previous analysis and definitions, we choose the Lyapunov functions of system \eqref{eq:CPD.b.2} as $ V: = V_{1}+V_{2}+\chi_1V_{3}$ with $V_3$ defined in Appendix \ref{app:thm-CPD.a} and satisfying \eqref{eq:CPD.a.V3}. Then we have
\begin{equation}
\ba \label{eq:CPD.b.Vp}
V\geq \frac{1}{2}\|\xchc\|^2+\frac{1}{2}\|\zb\|^2+ \frac{\chi_1}{2}\|\xpic\|^2+c_1\|\xl-\xhl\|.
\ea
\end{equation}
Combining \eqref{eq:CPD.b.V1}, \eqref{eq:CPD.b.V2}, \eqref{eq:CPD.a.V3}, and letting $r\leq1,\beta\leq\alpha,\eta\leq\beta^2_\alpha$, 
we can derive
\[\ba
    \dot{V}&\leq 
     -(\xi_{1} \alpha -\xi_{2} \beta_\alpha^2 - \xi_3 \alpha r)\|\xchc\|^2
     -(\beta^2_\alpha-\xi_4 \eta-\xi_5 \beta^2_\alpha r)\|\zb\|^2\\
    &\quad-(\eta \frac{\mu_n}{4}\chi_1) \|\xpic\|^2-(c_3-\xi_6 \alpha/r - \xi_7 \alpha)\|\xl-\xhl\|^2.
    \ea
\]
It can be noticed that $\dot{V}$ is negative definite when we choose $r=\mathrm{min}\left\{\frac{\xi_1}{4\xi_3},\frac{1}{4\xi_5},1\right\}$,  $\alpha\leq \mathrm{min}\left\{\frac{c_3r}{4\xi_6}, \frac{c_3}{4\xi_7}\right\}$, $\beta^2\leq \mathrm{min}\left\{{\alpha^2},\frac{\xi_1\alpha^2}{4\xi_2}\right\}$, $\eta\leq \mathrm{min}\left\{\beta_\alpha^2,\frac{1}{4\xi_4}\right\}$. With  \eqref{eq:CPD.b.Vp}, we have
 {
\[
\dot{V}\leq -\gamma V,\quad \gamma = \mathrm{min}\left\{\frac{\lambda_2\alpha}{2},\frac{\beta^2}{\alpha},\eta\frac{\mu}{2n},\frac{c_3}{2c_1}\right\}.
\]}
This yields  $\|\xl(t)\|^2=\mathcal{O}\left(e^{-\gamma t}\right)$ by the definition of $V(t)$. With the definition $\xl(t)=\xb(t)-\calSo$ before, we know that $\xb_i(t)$ in Flow \eqref{eq:CPD.b} converges exponentially to the optimal solution $s^\ast$  with the ST compressor. The proof is completed.

\section{Proof of Proposition \ref{pro-dis}} 

\emph{Proof of a).} By recalling \cite{Brian-TAC-1977}, it can be proved that ${\xb}_e\left(t+1\right)=\xb_e(t)-\kappa_0\psib(t)\psib(t)^T\xb_e(t)$ is UGLS at the zero equilibrium for any $\kappa_0\leq \kappa_0^\ast$ with some $\kappa_0^\ast>0$ under the condition in discrete-time cases. The remaining proof is similar to that in Proposition \ref{propos}.a) and is omitted here.

\emph{Proof of b).} 
Next, we prove that $\mathbf{C}_2$ is the ST compressor in discrete time by proving the system
   $\xb_e\left(t+1\right)=\xb_e(t)-\kappa_0\mathbf{C}_2\left(\xb_e(t)\right)$ is UGLS at the zero equilibrium with $\kappa_0=\frac{1}{p}$. By \eqref{ass_c2'}, there holds
\[
    \ba 
&\quad\|\xb_e\left(t+1\right)\|^2-\|\xb_e(t)\|^2\\&=-\frac{2\mathbf{C}_2\left(\xb_e(t)\right)^T\xb_e(t)}{p}+\left\|\frac{\mathbf{C}_2\left(\xb_e(t)\right)}{p}\right\|^2\leq -\varphi\|\xb_e(t)\|^2.
    \ea
\]
Thus $\xb_e$-system is UGLS at the zero equilibrium with $\varphi\in(0,1]$. With \eqref{eq:c2L}, the proof is complete.

 {
\emph{Proof of c).} 
Finally, we prove that $\mathbf{C}_3$ is the ST compressor in discrete time by proving the system $\xb_e\left(t+1\right)=\xb_e(t)-\kappa_0\gamma_e^t\left\lfloor\frac{\xb_e}{\gamma_e^t}\right\rfloor$
is UGLS at the zero equilibrium with $\kappa_0=1$. We let $\zb_e(t):=\frac{\xb_e(t)}{\gamma^t_e}$, then we have
\[
    \ba 
\zb_e\left(t+1\right)-\zb_e(t)&=-\frac{\left\lfloor\zb_e(t)\right\rfloor}{\gamma_e}+\frac{1-\gamma_e}{\gamma_e}\zb_e(t)
\\
\Rightarrow
\zb_e\left(t+1\right)&=  \frac{1-\gamma_e}{\gamma_e}(\zb_e(t)-\left\lfloor\zb_e(t)\right\rfloor),
    \ea
\]
which leads to $\|\zb_e(t)\|\leq \frac{1-\gamma_e}{\gamma_e}\sqrt{d}$ for any $t\geq0$. The remaining proof is the same as that in Proposition \ref{propos}.c) and we complete the proof.}

\section{Proof for Theorem 
\ref{thm-DPD.a}}
\label{app:thm-DPD.a}


Referring to the continuous-time flow in Appendix \ref{app:thm-CPD.a}, DPD-DC can be written in a compact form with the same equilibrium point as system \eqref{eq:CPD.a.1}.
Then we introduce the state error by defining $\overline{\xb}:=\xb-\calSo$, $\overline{\vb}:=\vb+\frac{\eta \Fb\left(\calSo\right)}{\beta}$,
and decompose $\xl$ and $\vl$ by defining $\xch:=\calSb^T\xl$, $\xpi:=\onb^T\xl$, $\vch:=\calSb^T\vl$ and  $\vpi:=\onb^T\vl$. The convergence of $\xl(t)$ and $\vl(t)$ follows by proving  {$\xpi(t)$, $\xch(t)$, $\vpi(t)$ and $\vch(t)$} converge to the zero equilibrium. In addition, we can conclude that \eqref{eq:Hv} holds.

Consider $\zb:=\frac{1}{\beta}\vch+\xch$, then DPD-DC is equal to the following system for $({\overline{\xb}}_{\perp}, {\overline{\xb}}_{\parallel}, {\zb})$, as
\begin{equation}
\label{eq:DPD.a.3}
\ba 
\mathbf{{\overline{x}}}_{\perp}\left(t+1\right)&=\mathbf{{\overline{x}}}_{\perp}(t)-\kappa_0\calSb^T \mathbf{L}_\otimes \xb_c(t)\\
&\quad +\kappa[\beta^2\xch(t)-\beta^2 \zb(t) -\eta\calSb^T \Fl\left(\xl(t)\right)],\\
\mathbf{{\overline{x}}}_{\parallel}\left(t+1\right)&=\mathbf{{\overline{x}}}_{\parallel}(t)-\kappa\eta\onb^T\Fl\left(\xl(t)\right),\\
    {\zb\left(t+1\right)}&=\zb(t)+\kappa[-\beta^2 \zb(t) +\beta^2\xch(t)- \eta\calSb^T \Fl\left(\xl(t)\right)],\\
\xb_c(t)&=\Cb \left(\xb(t),t\right)-\Cb\left(\onb\xpi(t)+\calSo,t\right) .
    \ea.
\end{equation}

 In the following, we  investigate the stability of system \eqref{eq:DPD.a.3}.
Let $V_{1,t}\left(\xch,\zb\right)=\frac{1}{2}\left(\|\xch\|^2+\|\zb\|^2\right)$, whose difference is given by
\begin{equation}
    \label{eq:DPD.a.V1}
\ba 
\Delta{V}_{1,t} 
&\leq \left(L_c\lambda_n\kappa_0+2L_c^2\lambda^2_n\kappa_0^2\right)\|\xch\|^2\\
&\quad +\kappa[-\left(\beta^2-\frac{\eta}{2}\right) \|\zb\|^2 +\left(\beta^2+\frac{\eta}{2}+{\eta}L_f^2\right) \|\xch\|^2
\\&\quad +\eta L_f^2\|\xpi\|^2]+\frac{1}{2}\kappa^2[\left(7\eta^2L_f^2+7\beta^4\right)\|\xch\|^2\\&\quad +7\beta^4\|\zb\|^2+7\eta^2L_f^2\|\xpi\|^2],
\ea
\end{equation}
where the inequality is obtained by \eqref{eq:Fl} and \eqref{eq:xc}.

Before we introduce the second Lyapunov function, we will show that the following system,
\begin{equation}
\label{eq:DPD.aux}
\ba 
{\yb}_{e}\left(t+1\right)=\yb_{e}(t)-\kappa_0\calSb^T\mathbf{L}_\otimes\mathbf{\Cb}\left(\xb_{e}(t),t\right),
\ea  
\end{equation}
where $\yb_e\in\mathbb{R}^{\left(n-1\right)d}$, $\xb_e\in\mathbb{R}^{nd}$ and ${\yb}_{e}=\calSb^T{\xb}_{e}$, achieves UGLS at the zero equilibrium for some $\kappa_0,\delta$. 

By P1') of $\mathbf{C}\left(\xb_e,t\right)$, it is easy to find the following system achieves UGLS at the zero equilibrium if  $\kappa_0\leq\kappa_0^\ast/\min\{\lambda_n,1\}$, 
\[
\ba 
{\yb}_{e}\left(t+1\right)=\yb_{e}(t)-\kappa_0\Lambda \overline{\Cb}\left(\yb_{e}(t),t\right).
\ea  
\]
Then there exist positive constants $C$, $\gamma_D<1$ such that for any $t$ and $N\in\mathbb{N}_+$, the solution satisfies
\[
\left(\|\yb_{e}\left(t+N\right)\|^2\right)\leq C\left(\|\yb_{e}(t)\|^2\right)\gamma_D^N.
\]
We assume $\phi_t^{t+T}\left(\yb_{e}(t)\right)$ is the state of the system $\yb_{e}\left(t+1\right)=\yb_{e}(t)-\kappa_0\Lambda \overline{\Cb}\left(\yb_{e}(t),t\right)$ in $t+T$ moment for any $0\leq T\leq N$ with the state in $t$ moment is $\yb_{e}(t)$.
It is easy to verify that there exists some $L_\phi>0$ that
$\|\phi_t^{t+T}\left(\yb\right)\|^2\leq L_\phi \|\yb\|^2$ holds
for any $\yb\in\mathbb{R}^{\left(n-1\right)d}$ and $0\leq T\leq N$  by P2') of the compressor $\mathbf{C}$.

We define a Lyapunov function $V_{e,t}\left(\yb_e,t\right):=\sum_{j=0}^{N-1}\|\phi_t^{t+j}\left(\yb_e\right)\|^2$ satisfying
\begin{equation}
    \label{eq:DPD.a.Ve}
    \ba     &\quad c_{1}\|\yb_{e}\|^2\leq V_{e,t}\leq   c_{2}\|\yb_e\|^2
    \ea
\end{equation}
for $c_{1}=1,c_{2}=NL_\phi$.

In addition, we have
\begin{equation}
    \label{eq:V_e1}
\ba 
    \Delta V_{e,t}&=\sum_{j=1}^{N}\|\phi_{t+1}^{t+j}\left(\yb_{e}\left(t+1\right)\right)\|^2-\sum_{j=0}^{N-1}\|\phi_t^{t+j}\left(\yb_e(t)\right)\|^2\\
    &=\|\yb_{e}\left(t+N\right)\|^2-\|\yb_e(t)\|^2\\
    &\leq -\left(1-C\gamma_D^{N}\right)\|\yb_e(t)\|^2
    \leq -c_3\lambda_2\hat{\kappa}\|\yb_e(t)\|^2
    \ea
\end{equation}
for $\hat\kappa:=\frac{\kappa_0}{\kappa_0^\ast}\leq\min\{\frac{1}{\lambda_n},1\}$. Notably, for convenience in the subsequent analysis and to highlight the effect of $\Lambda$, we use $c_3\lambda_2$, instead of a single parameter, in the middle inequality of \eqref{eq:V_e1}.

We choose a $N\in\mathbb{N}_+$ large enough and then $c_{3}:=\frac{1-C\gamma_D^{N}}{\lambda_2\hat{\kappa}}>0$, i.e.,
\begin{equation}
    \ba 
    \label{eq:DPD.a.Vec3}
&\quad \sum_{j=1}^{N}\|\phi_{t+1}^{t+j}\left(\yb_e-\hat{\kappa}\Lambda \overline{\Cb}\left(\yb_e,t\right)\right)\|^2-\sum_{j=0}^{N-1}\|\phi_t^{t+j}\left(\yb_e\right)\|^2\\&\leq -c_{3}\lambda_2\hat{\kappa}\|\yb_e\|^2.
    \ea
\end{equation}
In addition, we have 
\begin{equation}
\ba \label{eq:DPD.a.fact1}
\|\yb_e-\kappa_0\Lambda \overline{\Cb}\left(\yb_e,t\right)\|^2\leq \theta\|\yb_e\|^2,
\ea
\end{equation}
for $\theta:=2+2
L_c^2\kappa_0^2\lambda_n^2>0$ by P2') of $\mathbf{C}$.

For system \eqref{eq:DPD.aux}, we use the Lyapunov function $V_{e,t}\left(\yb_e,t\right)$ and obtain the difference of it, as
\begin{equation}
  \ba   \label{eq:DPD.a.Ve'}
&\quad \sum_{j=1}^{N}\|\phi_{t+1}^{t+j}\left(\yb_e-\kappa_0\calSb^T\mathbf{L}_\otimes\mathbf{\Cb}\left(\xb_e(t),t\right)\right)\|^2 -\sum_{j=0}^{N-1}\|\phi_t^{t+j}\left(\yb_e\right)\|^2\\&\leq -c_{3}\lambda_2\hat\kappa\|\yb_e\|^2+c_4\kappa_0^2\lambda_n^2\|\overline{\Cb}\left(\calSb^T\xb_e(t),t\right)-\calSb^T\Cb\left(\xb_e(t),t\right)\|^2\\&\quad +2c_4\kappa_0\lambda_n\|\yb_e\|\|\overline{\Cb}\left(\calSb^T\xb_e(t),t\right)-\calSb^T\Cb\left(\xb_e(t),t\right)\|\\&\leq-\left(c_3\lambda_2\hat{\kappa}-2c_4\kappa_0\lambda_n\delta-c_4\kappa_0^2\lambda_n^2\delta^2\right)\|\yb_e\|^2
  \ea  
\end{equation}
for $c_{4}:=NL_\phi\theta$, where the first inequality is obtained by \eqref{eq:DPD.a.Vec3} and the second inequality is obtained by \eqref{propery iii}. It is obvious that for 
$
0<\delta<\frac{c_4+\sqrt{c_4^2+c_3\lambda_2c_4\hat{\kappa}}}{c_4\kappa_0\lambda_n}
$, the difference of $V_e\left(\yb_e,t\right)$ is negative definite with $c_3':=c_3-\frac{2c_4\kappa_0\lambda_n\delta-c_4\kappa_0^2\lambda_n^2\delta^2}{\lambda_2\hat{\kappa}}>0$. Thus system \eqref{eq:DPD.aux} achieves UGLS at the zero equilibrium.

Next we continue to choose the Lyapunov function for system \eqref{eq:DPD.a.3} by letting $V_{2,t}(\xch,t)=V_e\left(\xch,t\right)$, then we have
\begin{equation}
\ba \label{eq:DPD.a.V2}
\Delta V_{2,t} &= \sum_{j=1}^{N}\|\phi_{t+1}^{t+j}
\left(\xch\left(t+1\right)\right)
\|^2-\sum_{j=0}^{N-1}\|\phi_t^{t+j}\left(\xch(t)\right)\|^2\\
&\leq -\left(\hat{\kappa}\lambda_2c'_{3}-\kappa c_4 \left(\beta^2-
\frac{\beta^2}{r}- \frac{\eta}{r}-\eta rL_f^2\right)\right)\|\xch\|^2\\&\quad+ \kappa\big(c_{4}\beta^2 r\|\zb\|^2+c_{4}\eta rL_f^2\|\xpi\|^2\big)
\\
&\quad +\kappa^2NL_\phi\big(\left(3\beta^4+3\eta^2L_f^2\right)\|\xch\|^2\\
&\quad +3\beta^4\|\zb\|^2+3\eta^2L_f^2\|\xpi\|^2\big),
\ea
\end{equation}
where $r>0$ is a parameter to be determined later, and the inequality is obtained by \eqref{eq:DPD.a.fact1}, \eqref{eq:DPD.a.Ve'}, \eqref{eq:Fl} and the Young's Inequality.

Let $V_{3,t}\left(\xpi\right)=\frac{1}{2}\|\xpi\|^2$. With \eqref{eq:CPD.a.V3} in mind, we have 
\begin{equation}
    \label{eq:DPD.a.V3}
    \ba 
    \Delta V_{3,t}&=\frac{1}{2}\left(\xpi-\kappa\eta\onb^T \Fl\left(\xl\right)\right)^T\left(\xpi-\kappa\eta\onb^T \Fl\left(\xl\right)\right)-\frac{1}{2}\xpi^T\xpi\\
    &\leq \kappa \left(-\eta\frac{\mu_n}{2} \|\xpi\|^2+\eta\frac{1}{2\mu_n}L_f^2 \|\xch\|^2 \right)\\
    &\quad +\frac{1}{2}\kappa^2\eta^2L_f^2\left(\|\xch\|^2+\|\xpi\|^2\right),
    \ea
\end{equation}
where the equality is obtained by \eqref{eq:Fl} and \eqref{eq:CPD.a.V3}.

For convenience of subsequent analysis, let us introduce some parameters 
$\chi_0$, $\chi_1$, $\xi_1$, $\xi_2,\dots>0$, which are independent of $\beta$, $r$ and $\eta$, and some parameters $\zeta_1$, $\zeta_2,\dots>0$, as
\[
\ba 
&\quad \chi_0=\left(2L_c\lambda_n\kappa_0+4L_c^2\lambda_n^2\kappa_0^2\right)/\lambda_2c'_3\hat\kappa,\\&\quad
\chi_1=\frac{4L_f^2}{\mu_n}+\frac{4\chi_0c_4L_f^2}{\mu_n},\\
&\quad \xi_1=\frac{3}{2}+L_f^2+\chi_0\left(c_4+c_4L_f^2\right)+\chi_1\frac{L_f^2}{2\mu_n},\\
&\quad \xi_2=2\chi_0c_4,\quad
\xi_3=\chi_0c_4,\quad
\xi_4=\frac{1}{2},\\
&\quad \zeta_1=\frac{7}{2}\eta^2L_f^2+\frac{7}{2}\beta^4+\chi_0NL_\phi\left(3\beta^4+3\eta^2L_f^2\right)+\frac{1}{2}\chi_1\eta^2L_f^2,\\
&\quad \zeta_2=\frac{7}{2}\beta^4+3\chi_0NL_\phi\beta^4,\\
&\quad \zeta_3=\frac{7}{2}\eta^2L_f^2+3\eta^2\chi_0NL_\phi L_f^2+\frac{1}{2}\eta^2\chi_1L_f^2.
\ea
\]

In view of the previous analysis and definitions, 
we choose the Lyapunov functions of system \eqref{eq:DPD.a.3} as $ V_t: = V_{1,t}+\chi_0V_{2,t}+\chi_1V_{3,t}$, which satisfies
\begin{equation}
\ba \label{eq:DPD.a.Vp}
V_t\geq \left(\frac{1}{2}+\chi_0c_1\right)\|\xch\|^2+\frac{1}{2}\|\zb\|^2+\frac{\chi_1}{2} \|\xpi\|^2.
\ea
\end{equation}
Combining \eqref{eq:DPD.a.V1}, \eqref{eq:DPD.a.V2}, \eqref{eq:DPD.a.V3}, and letting $r\leq1$, $\eta\leq\beta^2$, $\kappa\leq1$, we have
\[
    \ba
     \Delta V_t&\leq 
     -(\frac{1}{2}\chi_0c'_{3}-\kappa(\xi_1\beta^2+\xi_2\beta^2/r)-\kappa^2\zeta_1)\|\xch\|^2
     \\
    &\quad-(\kappa(\beta^2-\xi_3\beta^2r-\xi_4\eta)-\kappa^2\zeta_2)\|\zb\|_P^2\\
    &\quad-(\kappa\eta \frac{\mu_n}{4}\chi_1-\kappa^2\zeta_3) \|\xpi\|^2.
    \ea
\]
It can be noted that $\Delta V_t$ is negative definite when we choose $r=\mathrm{min}\left\{\frac{1}{4\xi_3},1\right\}$, $\beta^2\leq \mathrm{min}\left\{\frac{\chi_0c'_{3}\lambda_2\hat{\kappa}}{8\xi_1}, \frac{\chi_0c'_{3}r}{8\xi_2}\right\}$, $\eta\leq \mathrm{min}\left\{\beta^2,\frac{\beta^2}{4\xi_4}\right\}$ and $\kappa\leq \kappa_1:=\frac{1}{2}\mathrm{min}\left\{\frac{\chi_0c'_{3}\lambda_2\hat{\kappa}}{4\zeta_1},\frac{\beta^2}{2\zeta_2},\eta\frac{\mu_n\chi_1}{4\zeta_3},1\right\}$. With \eqref{eq:DPD.a.Vp} in mind, we have
 {
\begin{equation}
    \label{eq:rate_dc}
\ba 
    \Delta V_t \leq -\gamma V_t,\ \gamma =\frac{1}{2} \kappa\mathrm{min}\left\{\frac{L_c\lambda_nc_3'\lambda_2\kappa_0}{\lambda_2c_3'+4L_c\lambda_nc_1\kappa^\ast_0},\beta^2,\eta\frac{\mu}{2n}\right\}.
\ea
\end{equation}}
Let $\kappa_2:=2/\mathrm{min}\left\{\frac{\chi_0\lambda_2c'_3\hat{\kappa}}{2\left(1+2\chi_0c_1\right)},\beta^2,\eta\frac{\mu_n}{2}\right\}$. When $\kappa\leq\mathrm{min}\left\{\kappa_1,\kappa_2\right\}$, we can derive
for some $\gamma\in\left(0,1\right)$ , $V_t=\mathcal{O}\left(\left(1-\gamma\right)^t\right)$. It can be derived that $\|\xl(t)\|^2=\mathcal{O}\left(\left(1-\gamma\right)^t\right)$ by the definition of $V_t$. This implies by the definition $\xl(t)=\xb(t)-\calSo$  that $\xb_i(t)$ in DPD-OC converges linearly to the optimal solution $s^\ast$  with the SST compressor. The proof is completed.

\section{Proof of Theorem \ref{thm-DPD.b}}
\label{app:thm-DPD.b}
Using the same definitions of {$\xpi(t)$, $\xch(t)$, $\vpi(t)$ and $\vch(t)$} in Appendix \ref{app:thm-DPD.a}, with system \eqref{eq:CPD.b.2} in mind, we know DPD-OC is equal to the following system, as
\begin{equation}
\label{eq:DPD.b.1}
\ba 
\mathbf{{\overline{x}}}_{\perp}\left(t+1\right)&=\mathbf{{\overline{x}}}_{\perp}(t)-\kappa\calSb^T\mathbf{L}_\otimes\xhl(t)\\
&\quad +\kappa[\beta^2\xch(t)-\beta^2 \zb(t) -\eta\calSb^T \Fl\left(\xl(t)\right)],\\
\mathbf{{\overline{x}}}_{\parallel}\left(t+1\right)&=\mathbf{{\overline{x}}}_{\parallel}(t)-\kappa\eta\onb^T\Fl\left(\xl(t)\right),\\
    {\zb\left(t+1\right)}&=\zb(t)+\kappa[-\beta^2 \zb (t)+\beta^2\xch(t)- \eta\calSb^T \Fl\left(\xl(t)\right)],\\\xhl\left(t+1\right)&=\xhl(t)+\kappa_0\Cb\left(\xl(t)-\xhl(t),t\right).
    \ea
\end{equation}

Next, we will introduce some Lyapunov functions for system \eqref{eq:DPD.b.1}.
Let $V_{1,t}\left(\xch,\zb\right)=\frac{1}{2}\left(\|\xch\|^2+\|\zb\|^2\right)$, then we have
\begin{equation}
    \label{eq:DPD.b.V1}
\ba 
\Delta{V}_{1,t} 
&\leq \kappa\frac{1}{2}\lambda_n\|\xl-\xhl\|^2+2\kappa^2\lambda_n^2\|\xhl\|^2 +\kappa[-\left(\beta^2-\frac{\eta}{2}\right) \|\zb\|^2 \\
&\quad +\left(-\frac{1}{2}\lambda_2+\beta^2+\frac{\eta}{2}+{\eta}L_f^2\right) \|\xch\|^2
+\eta L_f^2\|\xpi\|^2]\\
&\quad +\frac{7}{2}\kappa^2[\left(\eta^2L_f^2+\beta^4\right)\|\xch\|^2+\beta^4\|\zb\|^2+\eta^2L_f^2\|\xpi\|^2],
\ea
\end{equation}
where the inequality is obtained by \eqref{eq:Fl} and \eqref{eq:xSLx}.

We conduct the following analysis to obtain the second Lyapunov function.
Now that $\xb_e\left(t+1\right)-\xb_e(t)=-\kappa_0\mathbf{C}\left(\xb_e(t),t\right)$, where $\xb_e\in\mathbb{R}^{d}$, achieves UGLS at the zero equilibrium by P1') of the compressor $\mathbf{C}_1$, then we know $\yb_e\left(t+1\right)-\yb_e(t)=-\kappa_0\Cb\left(\yb_e(t),t\right)$, where $\yb_e\in\mathbb{R}^{nd}$, also achieves UGLS at the zero equilibrium. A function $V_{e,t}\left(\yb_e,t\right)=\sum_{j=0}^{N-1}\|\phi_t^{t+j}\left(\yb_e\right)\|^2$ with same definition process as that in Appendix \ref{app:thm-DPD.a} can be obtained, which satisfies
\begin{equation}
    \label{eq:DPD.b.Ve}
    \ba 
    &\quad c_{1}\|\yb_e\|^2\leq V_{e,t}\leq
    c_{2}\|\yb_e\|^2,\\
    &\quad \sum_{j=1}^{N}\|\phi_{t+1}^{t+j}\left(\yb_e-\kappa_0\Cb\left(\yb_e,t\right)\right)\|^2-\sum_{j=0}^{N-1}\|\phi_t^{t+j}\left(\yb_e\right)\|^2\\&\leq -c_{3}\|\yb_e\|^2\\
    \ea
\end{equation}
for $c_{1}=1,c_{2}=NL_\phi,c_{3}>0$. As $V_{e,t}\geq0$, we have
\begin{equation}
    \label{eq:c3c1}
    c_3\leq c_1
\end{equation}
In addition, 
\begin{equation}
\ba \label{eq:DPD.b.fact1}
\|\xl-\xhl+\kappa_0\Cb\left(\xl-\xhl,t\right)\|^2\leq \theta\|\xl-\xhl\|^2
\ea
\end{equation}
for $\theta=2+2
L_c^2\kappa_0^2>0$ by P2') of $\mathbf{C}$.
Moreover, we can conclude that \eqref{eq:CPD.b.fact} holds.

Next we continue to choose the Lyapunov function by letting $V_{2,t}\left(\xl-\xhl,t\right)= V_{e,t}\left(\xl-\xhl,t\right)$, and we can derive
\begin{equation}
    \label{eq:DPD.b.V2}
    \ba 
    \Delta V_{2,t}
    &= \sum_{j=1}^{N}\|\phi_{t+1}^{t+j}\left(\xl\left(t+1\right)-\xhl\left(t+1\right)\right)\|^2\\
    &\quad -\sum_{j=0}^{N-1}\|\phi_t^{t+j}\left(\xl(t)-\xhl(t)\right)\|^2\\
    &\leq-\big(c_3-\kappa ( c_{4} r+2
c_{4} \beta^2/r+c_{4} \eta/r\\
&\quad+ 2c_{4}  r\lambda_n^2)\big) \|\xl-\xhl\|^2+\kappa ( c_{4} \beta^2 r\|\zb\|^2\\
&\quad +(2c_{4}  r\lambda_n^2+c_4\beta^2 r+c_{4} \eta rL_f^2)\|\xch\|^2+c_{4} \eta rL_f^2\|\xpi\|^2)\\
&\quad +\kappa^2NL_\phi \big(4\lambda_n^2\|\xhl\|^2+4\beta^4 \|\vl\|^2+(4\beta^4\\
&\quad +4\eta^2 L_f^2) \|\xch\|^2+4\eta^2 L_f^2 \|\xpi\|^2 \big)
    \ea
\end{equation}
for $c_{4}:=NL_\phi\theta$, where $r>0$ is a parameter to be determined later, the inequality is obtained by $\vl=\calSb\vch$, \eqref{eq:DPD.b.Ve}, \eqref{eq:DPD.b.fact1}, \eqref{eq:Fl}, \eqref{eq:CPD.b.fact} and the Young's Inequality.

For convenience of subsequent analysis, let us introduce some parameters  $\chi_1$, $\xi_1,\dots>0$, which are independent of  $\alpha$, $\beta$, $r$ and $\eta$, and some parameters $\zeta_1$, $\zeta_2,\dots>0$, as
\[
\ba 
&\quad \chi_1=\frac{4L_f^2}{\mu_n}+\frac{4c_4L_f^2}{\mu_n},\quad
\xi_1=\frac{\lambda_2}{2},\\
&\quad \xi_2=\frac{3}{2}+L_f^2+c_4+c_4L_f^2+\chi_1\frac{L_f^2}{2\mu_n},\quad\xi_3=2c_4 \lambda_n^2,\\
&\quad 
\xi_4=\frac{1}{2},\quad
\xi_5=c_4,\quad
\xi_6=4c_4,\quad\xi_7=\frac{\lambda_n}{2}+2c_4\lambda_n^2,\\
&\quad \zeta_1=\frac{7}{2}\eta^2L_f^2+\frac{7}{2}\beta^4+4\lambda_n^2\\
&\qquad \quad+NL_\phi\left(8\lambda_n^2+4\beta^4+4\eta^2L_f^2\right)+\frac{1}{2}\chi_1\eta^2L_f^2,\\
&\quad \zeta_2=\frac{7}{2}\beta^4+4NL_\phi\beta^4,\ \zeta_3=\frac{7}{2}\eta^2L_f^2+4NL_\phi\eta^2L_f^2+\frac{1}{2}\eta^2\chi_1L_f^2,\\
&\quad \zeta_4=4\lambda_n+8NL_\phi\lambda_n^2.
\ea
\]

In view of the previous analysis and definitions, we define the Lyapunov functions of system \eqref{eq:CPD.b.2} as $ V_t: = V_{1,t}+V_{2,t}+\chi_1V_{3,t}$ with $V_{3,t}$ defined in Appendix \ref{app:thm-DPD.a} and satisfying \eqref{eq:DPD.a.V3}. Then we have
\begin{equation}
\ba \label{eq:DPD.b.Vp}
V\geq \frac{1}{2}\|\xch\|^2+\frac{1}{2}\|\zb\|^2+ \frac{\chi_1}{2}\|\xpi\|^2+c_1\|\xl-\xhl\|.
\ea
\end{equation}

Combining \eqref{eq:DPD.b.V1}, 
\eqref{eq:DPD.b.V2},   \eqref{eq:DPD.a.V3}, and letting $r\leq1$, $\beta^2\leq1$, $\eta\leq\beta^2$, we have
\[
    \ba
     \Delta V_t&\leq 
     \kappa\big(-(\xi_{1}  -\xi_{2} \beta^2 - \xi_3  r)\|\xch\|^2\\
    &\quad-(\beta^2-\xi_4 \eta-\xi_5 \beta^2 r)\|\zb\|^2-\eta \frac{\mu_n}{4}\chi_1 \|\xpi\|^2\\
    &\quad-(c_3/\kappa-\xi_6 /r - \xi_7 )\big)\|\xl-\xhl\|^2\\
    &\quad+\kappa^2\big(\zeta_1 \|\xch\|^2 + \zeta_2 \|\vl\|^2_P+ \zeta_3 \|\xl-\xhl\|^2 + \zeta_4 \|\xpi\|^2 \big).
    \ea
\]
It can be noted that
$\Delta{V_t}$ is negative definite when we choose $r=\mathrm{min}\left\{\frac{\xi_1}{4\xi_3},\frac{1}{4\xi_5},1\right\}$,  $\kappa\leq \kappa_1= \frac{c_3}{2\xi_6r+2\xi_7}$, $\beta^2\leq \mathrm{min}\left\{{1},\frac{\xi_1}{4\xi_2}\right\}$, $\eta\leq \mathrm{min}\left\{\beta^2,\frac{1}{4\xi_4}\right\}$ and $\kappa\leq\kappa_2:=\frac{1}{2}\mathrm{min}\left\{\frac{\xi_{1} }{2\zeta_1}, \frac{\beta^2}{2\zeta_2}, \eta \frac{\chi_{1}\mu_n}{4\zeta_3},\frac{c_{3}}{2\zeta_4}\right\}$. With \eqref{eq:DPD.b.Vp} in mind, we have
 {
\begin{equation}
    \label{eq:rate_oc}
\ba 
    \Delta V_t \leq -\gamma V_t,\ \gamma = \frac{1}{2}\kappa\mathrm{min}\left\{\frac{\lambda_2}{2}, {\beta^2}, \eta \frac{\mu}{2n},\frac{c_3}{2c_1}\right\}.
\ea
\end{equation}}
Let $\kappa_3:=2/\mathrm{min}\left\{{\xi_{1}}, {\beta^2}, \eta \frac{\mu_n}{2},\frac{c_3}{2c_1}\right\}$. When $\kappa\leq\mathrm{min}\left\{\kappa_1,\kappa_2,\kappa_3\right\}$, we can derive
for some $\gamma\in\left(0,1\right)$ , $V_t=\mathcal{O}\left(\left(1-\gamma\right)^t\right)$ and thus $\|\xl(t)\|^2=\mathcal{O}\left(\left(1-\gamma\right)^t\right)$. With the previous definition $\xl(t)=\xb(t)-\calSo$, we conclude that $\xb_i(t)$ in DPD-OC converges linearly to the optimal solution $s^\ast$  with the ST compressor. The proof is completed.

\section{The expression of 
$\gamma_{OC}$ and 
$\gamma_{DC}$ for compressor $\mathbf{C}_{1a}$.}
\label{app:c_3same}

In this section, we prove that the expressions of $\gamma_{DC}$ and $\gamma_{OC}$ are given by
$\gamma_{DC}=\frac{1}{2} \kappa\mathrm{min}\left\{\frac{c_3\lambda_2}{4c_1\lambda_n},\frac{c_3\lambda_2}{4c_1},\beta^2,\eta\frac{\mu}{2n}\right\}$ and $\gamma_{OC}= \frac{1}{2}\kappa\mathrm{min}\left\{\frac{\lambda_2}{2}, {\beta^2}, \eta \frac{\mu}{2n},\frac{c_3}{2c_1}\right\}
$. As $\delta=0$ when $\mathbf{C}_{1a}$ is used, the expressions can be easily obtained by \eqref{eq:rate_dc} and \eqref{eq:rate_oc} as long as we can prove the parameters $c_3,c_1$ in \eqref{eq:rate_dc} and \eqref{eq:rate_oc} are the same.

To distinguish, we will refer to 
$c_3,c_1$ in 
$\gamma_{OC}$ as 
$c_{3,OC},c_{1,OC}$ and 
$c_3$ in 
$\gamma_{DC}$ as 
$c_{3,DC},c_{1,DC}$. Next, we will prove that for compressor $\mathbf{C}_{1a}$, there holds $c_{3,OC}=c_{3,DC},c_{1,OC}=c_{1,DC}$.
For the following system in Appendix \ref{app:thm-DPD.a},
\[
\ba 
{\yb}_{e}\left(t+1\right)=\yb_{e}(t)-\kappa_0\Lambda \Cb\left(\yb_{e}(t),t\right),
\ea  
\]
noting that $\mathbf{C}_{1a}=\psib(t)\psib^T(t)\xb$, where $\psib(t)= \mathbf{e}_i $ with $i=1+\left(t\   \mathrm{mod}\ d\right)$ for $t\in\mathbb{N}$, we define $V_{e,t}:=\sum_{i=1}^{d}\|\phi_t^{t+j}\left(\xb_e\right)\|^2$, where $\phi_t^{t+j}$ is defined in Appendix \ref{app:thm-DPD.a}, then we have
$\|\yb_{e}\|^2\leq V_{e,t}\leq d\|\yb_{e}\|^2$
and $\Delta V_{e,t}
    \leq -\lambda_2\kappa_0\alpha_1\|\yb_e(t)\|^2.
$ Comparing the results with \eqref{eq:DPD.a.Vec3}, we have $c_{3,DC}=\kappa_0\alpha_1$ and $c_{1,DC}=1$.

For the following system in Appendix \ref{app:thm-DPD.b},
\[
\ba 
{\yb}_{e}\left(t+1\right)=\yb_{e}(t)-\kappa_0^\ast \Cb\left(\yb_{e}(t),t\right),
\ea  
\]
we define $V_{e,t}=\sum_{i=1}^n\|\phi_t^{t+j}\left(\xb_e\right)\|^2$, where $\phi_t^{t+j}$ is defined in Appendix \ref{app:thm-DPD.b}, then we have
$\|\yb_{e}\|^2\leq V_{e,t}\leq d\|\yb_{e}\|^2$
and $\Delta V_{e,t}
    \leq -\kappa_0\alpha_1\|\yb_e(t)\|^2
$ by $\kappa_0\leq\kappa_0^\ast$. Comparing the results with \eqref{eq:DPD.b.Ve}, we have $c_{3,OC}=\kappa_0\alpha_1$ and $c_{1,OC}=1$. Then we complete the proof.

\section{Proof of Theorem \ref{thm-DPD.d}}

The idea of proof is quite similar to that in Appendix \ref{app:thm-DPD.b}. We just recalculate $\Delta V_{2,t}$
with stochastic impact while the other proof process is the same.

Now that $\xb_e(t+1)-\xb_e(t)=-\kappa_0\mathbf{C}(\xb_e(t),t)$, where $\xb_e(t)\in\mathbb{R}^{d}$, achieves mean square exponential convergence at the zero equilibrium, then clearly $\yb_e(t+1)-\yb_e=-\kappa_0\Cb(\yb_e,t)$, where $\yb_e\in\mathbb{R}^{nd}$, achieves  also.
Then there exists positive constants $C$, $\gamma<1$, for any $t$ and $T\in\mathbb{N}_+$, the solution satisfies
\[
\Eb\|\yb_{e}(t+T)\|^2\leq C(\|\yb_e(t)\|^2)\gamma^T.
\]

Assume $\phi_t^{t+N}(\yb_e)$ is the state of system $\yb_e(t+1)-\yb_e(t)=-\kappa_0\Cb(\yb_e(t),t)$ in $t+N$ moment with the state in $t$ moment is $\yb_e(t)$.
It is easy to verified that 
\[\ba
\Eb\|\phi_t^{t+N}(\yb)\|^2&\leq& L_\phi \|\yb\|^2
\ea
\]
for any $\yb\in\mathbb{R}^{(n-1)d}$ and some $L_\phi>0$ by property of compressor $\mathbf{C}$.

With \eqref{eq:DPD.a.Vec3} in mind, we can proof Lyapunov function $V_0(\yb_e,t)=\sum_{j=0}^{N-1}\|\phi_t^{t+j}(\yb_e)\|^2$ with some $N>0$ satisfies 
\begin{equation}
    \label{eq:DPD.d.Ve}
    \ba
    &c_{1}\|\yb_e\|^2\leq\Eb (V_{e,t})\leq
    c_{2}\|\yb_e\|^2\\
    &\Eb\sum_{j=1}^{N}\|\phi_{t+1}^{t+j}(\yb_e-\kappa_0C(\yb_e,t))\|^2-\Eb\sum_{j=0}^{N-1}\|\phi_t^{t+j}(\yb_e)\|^2\\
    &\leq -c_{3}\|\yb_e\|^2\\
    \ea
\end{equation}
for $c_{1}=1,c_{2}=NL_\phi,c_{3}>0$. 

Besides, 
\begin{equation}
\ba\label{eq:DPD.d.fact1}
\Eb\|\xl-\xhl+\kappa_0\Cb(\xl-\xhl,t)\|^2\leq \theta\|\xl-\xhl\|^2,
\ea
\end{equation}
for $\theta=2+2
L^2_c\kappa_0^2>0$ by property  of $\mathbf{C}$.

Define $V_2(\xl-\xhl,t):= V_e(\xl-\xhl,t)$, with \eqref{eq:DPD.b.V2} in mind, we can derive
\begin{equation}
    \label{eq:DPDS.b.V2}
    \ba
    \Eb \Delta V_{2,t}&= \Eb\sum_{j=1}^{N}\|\phi_{t+1}^{t+j}(\xl(t+1)-\xhl(t+1))\|^2\\
    &-\Eb\sum_{j=0}^{N-1}\|\phi_t^{t+j}(\xl(t)-\xhl(t))\|^2)\\
    &\leq-[c_3-\kappa\big( c_{4} /r+2
c_{4} \beta^2/r+c_{4} \eta/r) \|\xl-\xhl\|^2] \\
&+\kappa\big( c_{4}  r\lambda_n^2\|\xhl\|^2+c_{4} \beta^2 r\|\zb\|^2+c_4\beta^2 r\|\xch\|^2+\\
&c_{4} \eta rL_f^2\|\xch\|^2+c_{4} \eta rL_f^2\|\xpi\|^2\big)\\
&+\kappa^2NL_\phi\big(4\lambda_n^2\|\xhl\|^2+4\beta^4 \|\vl\|^2+(4\beta^4\\
&+4\eta^2 L_f^2) \|\xch\|^2+4\eta^2 L_f^2 \|\xpi\|^2\big),
    \ea
\end{equation}
for $c_{4}:=NL_\phi\theta$, where the first inequality is obtained \eqref{eq:DPD.d.Ve} and \eqref{eq:DPD.d.fact1}, and $r>0$ is a parameter which will be determined later.

We define the same $V_t$ as that in Appendix \ref{app:thm-DPD.b}, then \eqref{eq:DPD.b.Vp} holds and we have
\[
\ba
    \Eb\Delta V_t \leq -\gamma V_t,\ \gamma = \frac{1}{2}\kappa\mathrm{min}\{\frac{\lambda_{2}}{2}, {\beta^2}, \eta \frac{\mu_n}{2},\frac{c_3}{2c_1}\}.
\ea
\]
with the same parameters. Then we can derive $\Eb\|\xl(t)\|^2=\mathcal{O}((1-\gamma)^t)$. With the definition $\xl=\xb-\calSo$ before, we know that the mean square of $\xb_i(t)$ in DPD-OC converge exponentially to the optimal solution $s^\ast$ with the StST compressor.

\end{document}